\documentclass[twocolumn,pra,aps,showpacs,superscriptaddress,floatfix]{revtex4-1}
\usepackage{graphicx,epstopdf}
\usepackage{dcolumn,slashed}
\usepackage{amsmath,amssymb}
\usepackage{bm}
\usepackage{hyperref}
\def\<{\left<}
\def\>{\right>}
\def\ket|#1>{\left|#1\right>}
\def\bra<#1|{\left<#1\right|}
\def\elem<#1|#2|#3>{\left<#1\right|#2\left|#3\right>}
\def\({\left(}
\def\){\right)}

\def\pl{\partial}
\def\Tr{\hbox{Tr}}
\def\d{{\rm d}}
\def\e{{\rm e}}

\begin{document}

\title[Short Title]{Synthetic Unruh effect in cold atoms}

\author{Javier Rodr\'{\i}guez-Laguna}
\affiliation{Dto. F\'{\i}sica Fundamental, Universidad Nacional de
  Educaci\'on a Distancia (UNED), Madrid, Spain}
\affiliation{ICFO-Institut de Ci\`{e}ncies Fot\`{o}niques, The
Barcelona
  Institute of Science and Technology, 08860 Castelldefels
  (Barcelona), Spain}

\author{Leticia Tarruell}
\affiliation{ICFO-Institut de Ci\`{e}ncies Fot\`{o}niques, The
Barcelona
  Institute of Science and Technology, 08860 Castelldefels
  (Barcelona), Spain}

\author{Maciej Lewenstein}
\affiliation{ICFO-Institut de Ci\`{e}ncies Fot\`{o}niques, The
Barcelona
  Institute of Science and Technology, 08860 Castelldefels
  (Barcelona), Spain}
\affiliation{ICREA-Instituci\'o Catalana de Recerca i Estudis
Avan\c{c}ats, Lluis Companys 23, 08010 Barcelona, Spain}

\author{Alessio Celi}
\affiliation{ICFO-Institut de Ci\`{e}ncies Fot\`{o}niques, The
Barcelona
  Institute of Science and Technology, 08860 Castelldefels
  (Barcelona), Spain}

%\date{August 1, 2016}

\begin{abstract}
We propose to simulate a Dirac field near an event horizon using
ultracold atoms in an optical lattice. Such a quantum simulator allows
for the observation of the celebrated Unruh effect. Our proposal
involves three stages: (1) preparation of the ground state of a
massless 2D Dirac field in Minkowski spacetime; (2) quench of the
optical lattice setup to simulate how an accelerated observer would
view that state; (3) measurement of the local quantum fluctuation
spectra by one-particle excitation spectroscopy in order to simulate a
De Witt detector. According to Unruh's prediction, fluctuations
measured in such a way must be thermal. Moreover, following Takagi's
inversion theorem, they will obey the {\em Bose-Einstein}
distribution, which will smoothly transform into the {\em Fermi-Dirac}
as one of the dimensions of the lattice is reduced.
\end{abstract}


\maketitle

%%%%%%%%%%%%%%%%%%%%%%%%%%%%%%%%%%%%%%%%%%%%%%%%%%%%%%%%%%%%%%%%%%%

\section{Introduction}

The path towards quantum gravity opened a territory full of surprises:
quantum field theory in curved spacetime
\cite{Birrell_Davies}. Bekenstein's phenomenological thermodynamics of
black holes \cite{Bekenstein.73} received a strong support from
Hawking, when he found that a black hole must emit thermal radiation
\cite{Hawking.75}. The discovery hinted that thermal effects might
appear without any underlying stochasticity. Fulling, Davies, and
Unruh proposed that a similar effect existed in an essentially flat
spacetime, i.e., Rindler spacetime: an accelerated observer through an
empty Minkowski spacetime will perceive a thermal bath of particles,
at a temperature proportional to its acceleration
\cite{Fulling.72,Davies.75,Unruh.76}. Both phenomena are intimately
related: in both cases, an event horizon, which prevents communication
between different regions of spacetime, is developed. Furthermore, in
order to observe Hawking's radiation one must stay at rest near a
black hole, and therefore feel an acceleration. A further surprise was
revealed when Takagi studied the relation between dimensionality and
the Unruh thermal spectrum \cite{Takagi.86}. In 3+1D, an accelerated
detector of bosonic particles in Minkowski spacetime will record a
Bose-Einstein distribution, and a detector of fermionic particles will
find a Fermi-Dirac distribution. But this is only true if the
dimension of space is odd. Otherwise, an apparent {\em statistics
  inversion} phenomenon takes place: bosons are detected with a
Fermi-Dirac distribution, while fermions are detected with a
Bose-Einstein distribution. The Unruh effect is not just an exotic
curiosity: it bears a deep relation to entanglement
\cite{Susskind_Lindesay} and black hole thermodynamics, and it
plays a central role in Jacobson's derivation of Einstein
equations as equations of state for spacetimes in thermal
equilibrium \cite{Jacobson.PRL.95}. These results point to a
fundamental nature of the Unruh effect as a quantum counterpart of
the principle of equivalence, which it corrects
\cite{Singleton.PRL.11}. Moreover, the Unruh effect can be
regarded as a particular case of {\em parametric
  amplification} of the vacuum fluctuations \cite{Nation.RMP.12},
which puts it in the same class of phenomena as the dynamical
Casimir effect \cite{Davies.77,Calogeracos.02a,Calogeracos.02b}.
The latter can be seen as a flat spacetime analog of the Hawking
effect and connected with the Unruh thermal bath close to the
black hole horizon. The intriguing relation between the Unruh and
dynamical Casimir effects has been also explored in the context of
brane physics \cite{Russo.CQG.08,Chernicoff.11}.

The fundamental relevance of the Unruh effect provides a strong
motivation to measure it and related phenomena in the laboratory
\cite{Crispino.RMP.08} (see also \cite{Fuentes.11} for some more
recent proposals). Given the difficulty of the task, a different
approach has been to develop analogue gravity systems where
Hawking radiation might show up \cite{Volovik.03,Barcelo.05}. One
of the first ideas \cite{Unruh.81} was to build a {\em sonic
analogue} of a black hole in a moving medium, whose speed of sound
replaces the speed of light. If the relative velocity between
parts of the propagating medium is larger than the speed of sound,
an effective horizon appears. The medium can be either water
\cite{Weinfurtner.PRL.11,Weinfurtner.13} or a Bose-Einstein
condensate (BEC)
\cite{Garay.PRA.01,Fedichev.03,Fedichev.04,Steinhauer.14,Steinhauer.16}, which
can be employed also to probe the dynamical Casimir effect
\cite{Fedichev.04b,Carusotto.10,Westbrook.12,Westbrook.15}. A specific proposal
for measuring the Unruh effect in this setting, using an
accelerated impurity as De Witt detector, was proposed in
\cite{Retzker.PRL.08} (impurities can be used also as detectors of
Casimir forces and quantum friction \cite{Marino.16}).
% Leticia response to referee 2-> clarify citation of Retzker paper
Other very interesting approaches are to use a non-linear optical
medium in which a refractive index perturbation moves at high
speed
\cite{Philbin.08,Belgiorno.PRL.10,Unruh.PRL.11,Unruh.PRD.12,Finazzi.14},
or to exploit the geometric properties of graphene sheets
\cite{Iorio.PLB.12,Cvetic.12,Iorio.14}. The use of engineered
lattices of superconducting qubits \cite{Nation.RMP.12} has been
already used to probe the dynamical Casimir effect
\cite{Wilson.Nat.11} and proposed for Unruh physics.

In this work, we take a different strategy and propose a new
framework for simulating the Unruh effect, which is based on the
{\it quantum simulation} of Dirac fermions using ultracold
fermionic atoms in a 2D optical lattice
\cite{Tarruell.12,Lewenstein.12}. The possibility of simulating
the Dirac Hamiltonian in certain spacetime metrics was recently
put forward by some of us, where the information about the metric
is encoded in the tunneling terms shaped by the lasers
\cite{Boada.11}. Building upon that framework, we propose to start
the experiment by setting up an optical lattice whose dynamics
simulates the massless Dirac Hamiltonian in 2+1D in Minkowski
spacetime, where the Fermi velocity, analogue to the speed of
sound in a BEC, plays the role of the speed of light. By achieving
the ground state, we can assume that our quantum state is the
Dirac vacuum in Minkowski spacetime. Now, we can {\em quench} the
system by suddenly changing the tunneling terms in the lattice to
the values corresponding to the Dirac Hamiltonian in a Rindler
metric. In other terms, the same Dirac physics --Minkowski vacuum
of Dirac fermions-- but viewed by an accelerated observer.
Canonical observation of the Unruh effect should be performed now
by a local {\em De Witt detector} \cite{Birrell_Davies}, a device
whose purpose is to couple minimally to the quantum fluctuations
of the field and interchange energy \cite{Takagi.86}. The full
spectrum of local fluctuations obtained is predicted to follow
both Unruh and Takagi's predictions.

What is the novelty of our approach? Our setup is a quantum simulator,
i.e., a {\em quantum computer of special purpose} \cite{Lewenstein.12}
that allows for a systematic study of gravitating quantum matter. For
instance, within our quantum simulator it is possible to change the
Fermi velocity or the shape of the metric. More importantly, it
provides a framework for systematically studying quantum many-body
systems \cite{Bloch.08}.  Beyond the free fields studied in this work,
let us emphasize that the setup we propose for simulating the Unruh
effect can be used also for studying interacting fermions in curved
spacetime, and that it allows for the subtle manipulations needed to
simulate experiments in relativistic quantum information
\cite{Alsing.12}. Another parameter which can be easily tuned is the
dimensionality of the artificial spacetime, thus allowing us to probe
the aforementioned inversion theorem of Takagi \cite{Takagi.86}.

The investigation of the Unruh effect bears a strong relation to the
study of boundary effects. Indeed, the horizon can be considered as a
boundary for fields which are accessible to the accelerated
observer. Ensuring that the boundary conditions do not spoil the
unitarity of the theory imposes certain conditions on the Hamiltonian
\cite{Ibort.05} which, as we will show, are fulfilled naturally for
the Dirac Hamiltonian in Rindler spacetime, and provides a procedure
to perform the right discretization. Surprisingly, our Hamiltonian has
the same form as one of the candidates to solve the Riemann conjecture
via the Hilbert-Polya approach, $H=xp$
\cite{Schumayer.RMP.11,Sierra.05,Sierra.11,Gupta.13}. In a different
line, our model bears relation to the hyperbolically deformed
Hamiltonians \cite{Ueda.09,Vekic.93} and to the techniques of
off-diagonal confinement in optical lattices \cite{Rousseau.10}.

As our work is meant to be directed to a wide audience, we try as much
as possible to keep it self-contained. In section \ref{sec:rindler} we
provide a pedagogical overview of the relativistic physics for an
accelerated observer, both classical and quantum. Section
\ref{sec:dirac} introduces the Dirac Hamiltonian in Rindler spacetime
and discusses its discretization.  Readers mainly interested in the
proposed quantum simulation of the Unruh effect could go directly to
section \ref{sec:proposal}, where we detail our quench strategy,
provide numerical simulations of the expected results and suggest a
possible experimental implementation. We finish in section
\ref{sec:conclusions} with conclusions and proposals for further work.

%%%%%%%%%%%%%%%%%%%%%%%%%%%%%%%%%%%%%%%%%%%%%%%%%%%%%%%%%%%%%%%%%%%%%%%%%

\section{Review of Rindler spacetime and QFT in curved spacetime}
\label{sec:rindler}

This section is a review of the physics of an accelerated
observer. We will discuss in a pedagogical fashion the basics of
the Rindler metric, the thermalization theorem and the Unruh
effect.

\subsection{Rindler spacetime}

\begin{figure}
\includegraphics[width=9cm]{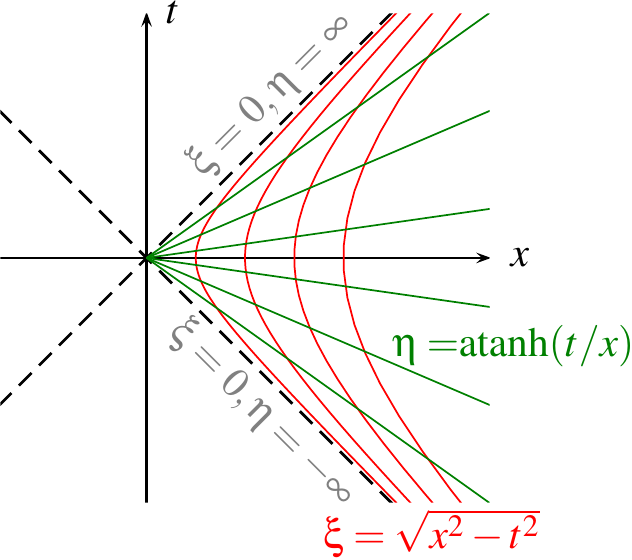}
\caption{Rindler coordinates on 1+1D Minkowski spacetime, $\eta$
  (Rindler time) and $\xi$ (Rindler space). The Rindler wedge,
  delimited by dashed lines, is the domain of validity of the
  coordinate patch. Constant $\eta$ lines (green) are spacelike, and
  constant $\xi$ lines (red) are timelike. For simplicity, we plot the
  trajectories only in the right wedge, $x>0$, as the ones for the
  wedge $x<0$ can be obtained by reflection around the $y$-axis. Note
  that as detailed in Sect. \ref{sec:proposal} both wedges are
  realized and are at the same footing in our proposal.  }
\label{fig:rindler}
\end{figure}

Let us briefly review Rindler physics, i.e., Minkowski spacetime
viewed by an accelerated observer \cite{Misner,Wald,Sachs}. Let us
consider an observer moving with constant acceleration $a=1$ (for
convenience in the following we take the speed of light to be $c=1$)
in the positive $x$-axis, at rest at $t=0$ and $x=1$. Physics seen by
this observer is more properly described in a co-moving reference
frame, obtained by the {\em Fermi-Walker} transport procedure. Let
$\eta$ be the co-moving time coordinate for this observer, and $\xi$
the co-moving space coordinate. They are called {\em Rindler
  coordinates}, and can be found using this transformation (see
Fig. \ref{fig:rindler})
\begin{equation}
\begin{cases}
t=\xi \sinh \eta\\
x=\xi \cosh \eta
\end{cases}.
\label{rindler.coord}
\end{equation}
In particular, the considered trajectory corresponds to $\xi=1$ for
all $\eta$. Notice the similarity with polar coordinates, where $\xi$
plays the role of a radius and $\eta$ is an angle in hyperbolic
geometry. The principle of equivalence states that physics seen by a
non-inertial observer can be absorbed by a change in her
metric. Indeed, in these coordinates, the Minkowski metric becomes
\begin{equation}
ds^2 = - \xi^2 d\eta^2 + d\xi^2 + dy^2 + dz^2,
\label{rindler.metric}
\end{equation}
which is known as the Rindler metric. Notice that the Rindler time
direction corresponds to a symmetry of the metric, i.e., it
constitutes a Killing vector which is inequivalent to the usual
Minkowski time direction. In fact, it corresponds to a boost
transformation. In the polar coordinates view, it is the generator of
hyperbolic rotations. Furthermore, the pole $\xi=0$ corresponds to a
singularity in the coordinate system, because the coefficient of
$d\eta^2$ vanishes. This is the hallmark of an {\em event
  horizon}. In fact, one can consider the Rindler metric
\eqref{rindler.metric} as a particular example of {\em optical
  metric} where the only non-trivial entry of the metric is
$g_{00}$, which becomes position-dependent. These are called
optical metrics because propagation of light along the geodesics
is equivalent to ``Galilean'' propagation in a refractive medium
with a position-dependent refractive index $n(x)$, which implies
the ``local'' speed of light $c_{loc}(x)=1/n(x)$. The
corresponding optical metric is of the form
\begin{equation}
ds^2 = - c_{loc}^2(x,y,z) dt^2 + dx^2 + dy^2 + dz^2.
\label{optical.metric}
\end{equation}

For Rindler spacetime, $c^2_{loc}(\xi)=\xi^2$. As $\xi\to 0$, the local speed of
light vanishes, which implies that signals cannot propagate beyond
that point. Thus, spacetime is separated into two parts which do not
communicate: the two Rindler wedges, $\xi>0$ and $\xi<0$. It is remarkable
that an event horizon can appear even in a flat spacetime.

Let us return to the proposed accelerated observer, which in Rindler
spacetime just sits at rest at $\xi=1$. From her point of view, light
moves at her left more slowly than usually, and faster at her
right. Near the horizon, $\xi=0$, light moves more and more slowly,
coming to stop at $\xi=0$, i.e., its local speed of light is zero
(while the actual speed of light stays obviously constant to 1). Let us now
consider objects which are static with respect to the accelerated
observer, i.e., objects at rest in Rindler spacetime at different
values of $\xi$. Tracing back their trajectories to Minkowski
spacetime, it can be checked that they correspond also to accelerated
trajectories, with acceleration $a(\xi)=1/\xi$. This implies that, in
order to keep pace with an observer with acceleration $a$ in front of
you, you must accelerate faster than that \cite{Wald}. This result is
known as {\em Bell's spaceship paradox}.

\subsection{The thermalization theorem}\label{sect:thermalization}

The interplay between thermodynamics and
general relativity gives raise to surprising properties.
Let us restrict ourselves to spacetimes which
contain a time-like Killing vector, i.e., spacetimes whose metric can
be said to be time invariant, where we have a well defined concept of
energy. In that case, {\em Tolman-Eherenfest's theorem}
\cite{Tolman.30,Takagi.86} states that for any field in thermal
equilibrium within a stationary curved spacetime, the product of the
local temperature and the modulus of the local time-like Killing
vector is constant, $T\cdot g_{00}^{1/2}=$ const. There is a simple
way to visualize this result. Photons emitted at one point $P$ in
space with frequency $\nu$ will reach another point in $P'$ with a
redshifted frequency, $\nu(P')=\nu(P)\cdot
\sqrt{g_{00}(P)/g_{00}(P')}$. Thus, the same factor should be applied
to energies and to temperatures. Thus, in Rindler spacetime,
temperature at any point is inversely proportional to the distance to
the horizon. Or, in other words, it is proportional to the
acceleration of an observer stationary at that point. Notice that this
does not entail a non-zero temperature. The theorem still holds if the
temperature is zero everywhere.

But the biggest surprises show up when we introduce quantum mechanics
\cite{Birrell_Davies}. Let us consider a free fermionic field in
Minkowski spacetime, with Hamiltonian $H_M$, described in terms of
local creation operators $c^\dagger_x$. The physical vacuum is the
ground state of its Hamiltonian, $\ket|0_M>$, and it does not
correspond to the Fock vacuum, $\ket|\Omega>$, which is defined by
$c_x\ket|\Omega>=0$ for all $x$. In the physical vacuum, all the
negative energy single-particle modes will be occupied

\begin{equation}
\ket|0_M>=\prod_{\omega_k^M<0} b^\dagger_k \ket|\Omega>,
\label{minkowski_vacuum}
\end{equation}
where $b^\dagger_k$ creates the $k$-th mode, and $\hbar \omega_k^M$ is its
energy. According to the usual convention, $b^\dagger_k$ creates a
particle if $\omega^M_k>0$ and an anti-particle if
$\omega^M_k<0$. Therefore, the physical vacuum is built by occupying
all the anti-particle states, and none of the particle ones.

Let us now consider an accelerated observer moving through this
vacuum. She will see physics displayed not on Minkowski spacetime,
but on a Rindler metric \eqref{rindler.metric}. Let $H_R$ be the
appropriate Hamiltonian operator, which is also a free fermionic
Hamiltonian. Its single-particle modes are known as the Rindler
modes, $d^\dagger_q$. They have energies $\hbar \omega_q^R$ and
are solutions to the wave equation in the Rindler metric (here the
index $q$ labels the eigenstates, but does not correspond to
momentum in the acceleration direction, since translational
invariance is broken). The ground state of $H_R$ is

\begin{equation}
\ket|0_R>=\prod_{\omega_q^R<0} d^\dagger_q \ket|\Omega>.
\label{rindler_vacuum}
\end{equation}
Again, the Rindler modes will qualify either as particles, if
$\omega^R_q>0$, or anti-particles if $\omega^R_q<0$. The pure state
$\ket|0_M>$ does not need to be an eigenstate of $H_R$, much less its
ground state. From the point of view of the accelerated observer, who
measures energies with $H_R$, $\ket|0_M>$ is not the true vacuum any
more. How does this state look like to her? It is crucial to realize
that the Rindler metric has a horizon, which separates space into two
parts which cannot communicate. Thus, she will not detect $\ket|0_M>$,
but the {\em reduced density matrix} which results of tracing out the
hidden part
\begin{equation}
\rho_R = \Tr_L \ket|0_M>\bra<0_M|,
\label{density_matrix}
\end{equation}
where $\Tr_L$ means a trace over the left-out degrees of freedom. A
reduced density matrix can always be formally written as a thermal
state
\begin{equation}
\rho_R = \exp(-H_E),
\label{entanglement_Hamiltonian}
\end{equation}
where $H_E$ is called the entanglement Hamiltonian
\cite{Li_Haldane.PRL.08,Susskind_Lindesay}. Since the Minkowski vacuum
\eqref{minkowski_vacuum} is a Slater determinant, we can use Wick's
theorem in reverse to prove that the entanglement Hamiltonian must be
a free fermionic Hamiltonian \cite{Peschel.03}. In other terms, the
accelerated observer will see a thermal state of free particles. The
Minkowski vacuum is invariant under Lorentz boosts, which correspond
to time translations in Rindler spacetime. For $\rho_R$, this property implies

\begin{equation}
0=\dot\rho_R= -\frac i{\hbar} [\rho_R,H_R].
\label{commutator}
\end{equation}
Thus, $[H_E,H_R]=0$, i.e., the entanglement Hamiltonian and the
Rindler Hamiltonian must commute. In fact, they can be
non-trivially proved to be proportional, and the constant of
proportionality can be read as an inverse temperature
\begin{equation}
\rho_R = \exp\left( - \frac{H_R} {k_B T_U} \right),
\label{thermalization}
\end{equation}
where
\begin{equation}
k_B T_U= \frac {\hbar a}{2\pi}. \label{unruh.temp}
\end{equation}
%Leticia for referee reply->T_U is the Unruh temperature
Here $T_U$ is known as the \emph{Unruh temperature} and this
result, which is far more general than the particular case studied
here, is the {\em thermalization theorem} \cite{Takagi.86}. Thus,
the Unruh temperature does not appear because of any underlying
stochasticity. The loss of information which gives rise to the
thermal effect is related to the presence of the horizon.
%%% Added explanation stationarity
An important consequence of thermalization theorem is that Minkowski vacuum 
{\it appears to be stationary to an accelerated observer}. Indeed, while 
Minkowski vacuum is clearly not an eigenstate of Dirac Hamiltonian in Rindler spacetime and, thus,
evolves non-trivially in Rindler time, {\it what an accelerated observer detects is 
invariant under such time evolution} as the thermal state is a diagonal density matrix in the
Rindler eigenbases. This observation is crucial in our proposal for simulating the Unruh effect with
ultracold atoms, see Sect. \ref{sec:proposal}.    

\subsection{The Unruh effect}\label{sec:Unruh_effect}

Let us consider the canonical transformation between Minkowski
($b^\dagger_k$) and Rindler modes ($d^\dagger_q$)

\begin{equation}
d^\dagger_q = \sum_k U_{qk} b^\dagger_k.
\label{bogoliubov.transf}
\end{equation}
This is a Bogoliubov transformation in disguise, because a
positive energy Rindler mode (particle) requires both positive and
negative Minkowski modes for its expansion (particle and
anti-particle). In fact, as the acceleration is position
dependent, the relevant Bogoliubov transformation has to be
defined locally. In the continuous limit, the global Bogoliubov
transformation \eqref{bogoliubov.transf} is even ill-defined as
the eigenstates are not normalizable. In order to properly define
it we have to consider normalized states \cite{Takagi.86}, for
instance wave packets centered around a generic point ${\bf r}$.
On physical terms, this means that we can associate a well-defined
acceleration to the Rindler wave-packet. For practical purposes,
the wave-packet normalization is equivalent to restricting the
scalar product of the unnormalized modes to a small region
$D_{{\bf r},\epsilon}$ such that $|{\bf r}' -{\bf r}| \le
\epsilon$ of space around ${\bf r}$. With this definition, the
occupation of each Rindler mode on the Minkowski ground state is
given by
\begin{equation}
n_{q, {\bf r}} \equiv \int_{D_{{\bf r}, \epsilon}} \bra<0_M|
d^\dagger_q \ket|{\bf r}'> \bra<{\bf r}'|  d_q \ket|0_M> =
\sum_{\omega^M_k<0} |\tilde U_{qk}({\bf r})|^2, \label{occupation.number}
\end{equation}
and the thermalization theorem ensures that

\begin{equation}
n_{q, {\bf r}} = \frac 1{\exp(\hbar \omega^R_q/k_B T_U({\bf r})) + 1
}, \label{occupation.thermal}
\end{equation}
with $k_B T_U({\bf r}) = \frac {\hbar}{2\pi x}$ according to
Eq. \eqref{unruh.temp}, where $x$ is the spatial distance of the point
$\mathbf{r}$ from the horizon.

But the {\em Unruh effect} goes beyond the thermalization theorem,
because it is defined {\em operationally}, in terms of what a
local observer can measure. The so-called {\em De Witt} detector
\cite{Birrell_Davies,Takagi.86} is a device carried along with the
observer, which couples minimally to the fermionic field at a
spatial point ${\bf r}$, and can emit and absorb particles. Under
a large variety of circumstances it can be proved that the
probability amplitude of absorption/emission is given solely by
the \emph{Wightman function}
\begin{equation}
G(t)\equiv \bra<0_M| \; c^\dagger_{x(t)}(t)\; c_{x(0)}(0) \; \ket|0_M>.
\label{detector.response}
\end{equation}
Here, $x(t)$ is the trajectory for the observer -- for simplicity
we consider trajectories parallel to the $x$-axis and we omit remaining
constant spatial coordinates -- and $c^\dagger_x(t)$ is the
creation operator for a fermion at event $(x,t)$.  The Fourier
transform of $G(t)$, $G(\omega)$, is the {\em detector response
function}, which should be experimentally accessible, as we will
discuss later.

The formula \eqref{detector.response} makes equal sense in Minkowski
or in Rindler spacetimes, if we are allowed to abuse notation and
let $x$ and $t$ denote the coordinates in both. In Rindler
spacetime, the trajectory of an accelerated observer will be just
a constant $x(t)=x_0$. Let us define two different basis changes,
from Rindler space-localized states to Rindler and Minkowski
modes, respectively. At time $t=0$, if $c^\dagger_x$ creates a
particle at point $x$, we have
\begin{align}
b^\dagger_k =& \sum_x M_{kx} c^\dagger_x \nonumber, \\
d^\dagger_q =& \sum_x R_{qx} c^\dagger_x,
\label{rindler.minkowski}
\end{align}
where the unitary matrices $M_{kx}$ and $R_{qx}$ are the
single-particle wave functions of Minkowski and Rindler modes,
respectively, and determine the unitary transformation $U_{qk}$ in
\eqref{bogoliubov.transf}, $U_{qk}=\sum_x R_{qx} \bar M_{kx}$. Here
and in the following by the bar we denote the complex conjugate of the
matrix elements.
So we get
\begin{eqnarray}
G_{x_0}(\omega) &\equiv& \int {\rm d} t\; \hbox{e}^{-i\omega t} \bra<0_M|
c^\dagger_{x_0}(t) c_{x_0}(0)
\ket|0_M> \label{detector.response.full} \\
&=& \sum_{q,q'} \delta(\omega-\omega^R_q) \bar R_{qx_0} R_{q'x_0}
\sum_{\omega^M_k<0} \bar U_{qk}  U_{q'k}. \nonumber
\end{eqnarray}
Thus, the detector response function is strongly dependent on the form
of the Rindler and Minkowski modes, through $U$ and $R$.

Going beyond the thermalization theorem, Unruh predicted that the
distribution function $G(\omega)$ will be {\em thermal}. But a
surprise is hiding behind Eq. \eqref{detector.response.full} due to
the spatial dependence of the Rindler and Minkowski modes. If the
dimension of space is {\em odd}, then the response function of a
fermionic field will follow the Fermi-Dirac distribution function, as
one would expect. But if the dimension of space is {\em even},
$G_{x_0}(\omega)$ will follow a Bose-Einstein distribution. The
opposite is true for a free bosonic field.
% Ale->reply to referee 2
This fact, known as Takagi's inversion theorem \cite{Takagi.86}
stems from dimensional effects in wave propagation, analogous to
those observed for light propagating radially. In odd dimension,
the Huygens principle holds, and a pointlike perturbation after a
time $t$ is concentrated in a spherical shell of radius $vt$,
where $v$ is the propagation velocity. In even dimension, however,
not all the scattered waves propagate at the same $v$, the
Huygens' principle does not hold and the perturbation becomes
radially spread with time.

Alternative physical meanings of the detector response function
\eqref{detector.response.full} are worth mentioning. The first is
a measure of quantum fluctuations: $G_{x_0}(\omega)$ is the power
spectrum of the quantum noise \cite{Candelas_Sciama.77}. The
second is related to the dynamical Casimir effect. Let us consider
a physical plane in space, whose interaction with our fermionic
field can be expressed as a Dirichlet boundary condition. Now let
us move this plane with constant acceleration $a$. Then, the
stress-energy tensor at any point will depend on its current
distance to the plane. In fact, it can be proved
\cite{Candelas_Deutsch.77,Takagi.86} that the limit $a\to\infty$
can be made meaningful, thus providing a well defined
stress-energy tensor for the Rindler vacuum, which induces quantum
fluctuations that are probed by expression
\eqref{detector.response.full}.

%%%%%%%%%%%%%%%%%%%%%%%%%%%%%%%%%%%%%%%%%%%%%%%%%%%%%%%%%%%%%%%%%%%%%%%%%%

\section{Dirac fermions in a Rindler lattice}
\label{sec:dirac}

In this section we describe the behavior of Dirac fermions in Rindler
spacetime for one and two spatial dimensions. In particular, we
explicitly construct the corresponding Hamiltonian in a square
lattice. Indeed, since the Rindler metric has a time-like Killing
vector field, we can use a Hamiltonian formalism and discretize it to
get a simple tunneling model. The resulting model bears a surprising
resemblance to the $xp$ Hamiltonian used in the Hilbert-Polya approach
to proving the Riemann conjecture.  This point is further detailed in
Appendix \ref{sec:1drindler}.

\subsection{The Dirac Hamiltonian in Rindler spacetime}
\label{sec:Rindler_Ham}

Let us consider a relativistic massless fermionic field in two
dimensions, governed by the Dirac equation in Minkowski spacetime

\begin{equation}
\gamma^a \pl_a \psi= 0,
\label{dirac.eq}
\end{equation}
where the $\gamma^a$ are a representation of the Clifford algebra,
$\{\gamma^a,\gamma^b\}=2\eta^{ab}$, where $\eta_{ab}=Diag(-1,1,1)$
is the Minkowski metric and $a,b=0,1,2$. In \eqref{dirac.eq}, as
well as in the rest of the section, sums over repeated indices are
left implicit according to Einstein's convention. As it stands,
the equation is manifestly Lorentz covariant. Let us shift to a
Hamiltonian view, which is more convenient for simulation. In
other words, we single out the time-derivative

\begin{equation}
i\pl_0\psi = {\cal H} \psi = -i\gamma_0 \gamma^j\pl_j \psi, \ \
j=1,2. \label{dirac.ham.minkowski}
\end{equation}

Let us make the following choice for the $\gamma_a$ matrices in two
dimensions, $-\gamma_2=\sigma_x$, $\gamma_1=\sigma_y$, $\gamma_0=i
\sigma_z$. We obtain

\begin{equation}
i\pl_t \psi= -i \( \pl_x\sigma_x + \pl_y\sigma_y \) \psi.
\label{dirac.Eq.mink}
\end{equation}

Equation \eqref{dirac.eq} can be formulated on a general (curved)
background metric $g_{\mu\nu}$ as well. For spinor systems it is very
convenient to introduce the {\em vielbein}, which is a set of vectors
defined on the tangent manifold, $e^a_\mu(x)$, such that
$g_{\mu\nu}(x)=e^a_\mu(x)e^b_\nu(x) \eta_{ab}$. The parallel transport
for the vielbein vectors defines the {\em spin-connection},
$w^{ab}_\mu$, and allows a compact expression for the covariant
derivative of a spinorial field \cite{Wald}

\begin{equation}
\pl_\mu \psi \to D_\mu \psi \equiv \( \pl_\mu + \frac 14 w^{ab}_\mu
\gamma_{ab}\) \psi,
\label{covariant.derivative}
\end{equation}
where $\gamma_{ab}\equiv \frac 12 [\gamma_a,\gamma_b]$. By making use
of it, the Dirac equation reads

\begin{equation}
\gamma^\mu D_\mu \psi=0,
\label{dirac.Eq.covariant}
\end{equation}
where the {\em curved} gamma matrices $\gamma_\mu$ are defined by
$\gamma_\mu = \gamma_a e^a_\mu$, and the curved indices
$\mu=t,x,y$ are lowered and raised by contracting with the metric
$g_{\mu\nu}$ and its inverse $g^{\mu\nu}$.  When we single out the
time-derivative, we obtain again a Schr\"odinger equation of the
form

\begin{equation}
i\pl_t \psi = -i \gamma_t \( \gamma^j\pl_j + \frac 14 \gamma^j
w^{ab}_j \gamma_{ab} +  \frac 14 \gamma^t w^{ab}_t \gamma_{ab}\) \psi,
\label{dirac.eq.curved}
\end{equation}
where $j=x,y$.

Let us now consider the specific case of the 2D Rindler metric
\eqref{rindler.metric} $ds^2=-x^2dt^2+dx^2+dy^2$. The only
non-vanishing element of the spin-connection is $w^{01}_t=x/|x|$. With
the aforementioned choice for the $\gamma_a$ matrices, we get

\begin{equation}
i\pl_t \psi= -i \( \(|x|\pl_x + \frac 12 \frac x{|x|}\)\sigma_x + |x|\pl_y\sigma_y
\) \psi.
\label{dirac.Eq.rindler}
\end{equation}
Thus, the Hamiltonian density becomes

\begin{equation}
{\cal H}_R=-i \( \(|x|\pl_x + \frac 12 \frac x{|x|}\)\sigma_x + |x|\pl_y\sigma_y
\),
\label{rindler.ham}
\end{equation}
which is the single-particle Rindler Hamiltonian. Its
second-quantized form is simply

\begin{equation}
H_R = \int \d x \d y \, \bar \psi^\dag {\cal H}_R \psi. \label{HR}
\end{equation}
The same expression can obviously be derived by taking the Legendre
transformation of the Dirac Lagrangian in Rindler spacetime.

In intuitive terms, the $|x|$ term is related to the volume form,
$\sqrt{-g}=|x|$. The $1/2$ term comes for the covariant derivative and
it is essential to ensure the hermiticity of $H_R$. Indeed, this
factor cancels the so-called {\em deficiency indices}
\cite{Reed_Simon_2,Ibort.05}, i.e., allows us to treat the horizon at
$x=0$ as a boundary, ensuring that any boundary condition can be
imposed while respecting self-adjointness of the Hamiltonian.

This property is more evident once \eqref{HR} is cast in symmetric
fashion, i.e., the spatial derivatives act symmetrically both on
$\psi$ and $\psi^\dag$,

\begin{align}
 H_R & = \frac 12 \int \d x\d y \, ({\cal H} \psi)^\dagger \psi
 + \frac 12 \int \d x\d y \, \psi^\dagger {\cal H} \psi  \cr
& = \frac i2 \int \d x\d y \, |x|\left((\partial_x \psi^\dag)
 \sigma_x \psi + (\partial_y \psi^\dag) \sigma_y \psi \right. \cr
 &\ \ \ \ \ \ \ \ \ \ \ \ \ \ \ \ \ \ \ \ \ \ \ \ \ \ \left.
- \psi^\dag \sigma_x \partial_x\psi - \psi^\dag \sigma_y
\partial_y\psi\right). \label{symHRcont}
\end{align}
In this form, the propagation in Rindler metric is sensitive only
to the overall scale factor which determines a Fermi velocity that
changes linearly along the $x$ direction.

It is worth noticing that the equivalent symmetric formulation of
single-particle Hamiltonian \eqref{rindler.ham} is

\begin{equation}
{\cal H}_R =  \sqrt{x} \slashed{p} \sqrt{x},
\label{diracEq.xpx}
\end{equation}
which is also manifestly Hermitian. It will be further discussed
in the Appendix \ref{sec:1drindler}, in relation with the Riemann
conjecture.

\subsection{Discretizing the Rindler Hamiltonian}

The Minkowski and Rindler Dirac Hamiltonians in one and two
spatial dimensions can be suitably discretized on the lattice
\cite{Boada.11}. As shown in detail in the Appendix
\ref{subsec:discr}, a convenient way of doing this is to consider
a 1D-chain or a 2D-square lattice with non-interacting spinless
fermions

\begin{equation}
H=-\sum_{\<{\bf r},{\bf r}'\>} t_{{\bf r}{\bf r}'} c^\dagger_{\bf r} c_{{\bf r}'} + {\rm H.c.},
\label{eq:discr_ff}
\end{equation}
where the sum runs over all pairs of nearest neighbors sites ${\bf
r},{\bf r}'$. Since the 1D-Dirac models can be realized as a slice
along $x$ (defined as the direction perpendicular to the Rindler
horizon) of the 2D-Dirac ones, we focus on the latter case.
Hamiltonian \eqref{eq:discr_ff} can represent the dynamics of each of
the chiral components of the Minkowski Dirac Hamiltonian if the
tunneling terms $t_{{\bf r}{\bf r}'}$ have all the same modulus and
the sum of their phases around each plaquette is $\pi$. This
corresponds to the well known $\pi$-flux Hamiltonian \cite{Kogut.75,Affleck.88,
  Lim.09}. All possible choices of phases respecting the $\pi$-flux
condition are equivalent, as they are related by gauge
transformations. We will focus on the one corresponding to the
symmetry gauge for the synthetic gauge field associated to the
phases. Precisely,
\begin{multline}
H_M = -\sum_{m,n} t_0 \left(e^{i \frac{\pi}2 (m-n)} c^\dagger_{m+1,n} \right. \\
   \left.+ e^{i \frac{\pi}2 (m-n)} c^\dagger_{m,n+1}\right) c_{m,n} + {\rm H.c.},
\label{pi.flux.Hamiltonian}
\end{multline}
where we adopt Cartesian coordinates to parametrize the lattice,
${\bf r} =(m\, d, n\, d)$ and denote with $d$ the lattice spacing.

The discretized version of the Rindler Dirac Hamiltonian
\eqref{symHRcont} can also be chosen to be of the form
\eqref{eq:discr_ff}, but with spatially modulated tunnelings,
$t_{{\bf r}{\bf r}'}$. Each tunneling rate has to be proportional
to the average $x$ coordinate of each link, which represents the
distance from the horizon. We place the horizon at $x=0$
accordingly to the coordinates chosen in \eqref{optical.metric}.
The tunneling phases have to satisfy the same $\pi$-flux condition
as for the Dirac Hamiltonian in Minkowski space. For the symmetric
gauge choice of \eqref{pi.flux.Hamiltonian} we have

\begin{multline}
H_R = -\sum_{m,n} t'_0 \left((m+\tfrac 12)\,
e^{i \frac{\pi}2 (m-n)} c^\dagger_{m+1,n} \right. \\\
 \left. + m \, e^{i \frac{\pi}2 (m-n)} c^\dagger_{m,n+1}\right) c_{m,n}
+ {\rm H.c.},
\label{Rindler.pi.flux.Hamiltonian}
\end{multline}
% Leticia reply to referee 2-> other Dirac systems
The numerical simulation and the experimental implementation in
optical lattices of the Hamiltonians \eqref{pi.flux.Hamiltonian}
and \eqref{Rindler.pi.flux.Hamiltonian} will be discussed in the
next section. We would like to remark here that in principle any
other lattice realization of the Dirac Hamiltonian like the ones
in bichromatic \cite{Salger.11}, hexagonal \cite{Soltan-Panahi.11,
Duca.15} and brick-wall lattices \cite{Tarruell.12}, which do not
involve artificial gauge fields, can be considered and deformed by
shaping the tunneling term to reproduce the Dirac Hamiltonian in
Rindler spacetime. Other artificial lattice Dirac systems such as
nano-patterned 2D electron gases, photonic crystals, micro-wave
lattices \cite{Polini.13} or polaritons \cite{Jacqmin.14} could
also be used. Since the Unruh effect is a single-particle and
purely kinematic effect, it could be studied using both bosonic
and fermionic systems. The latter offers a simple route to explore
the relativistic (linear dispersion relation) regime, as detailed
in the next section.

%%%%%%%%%%%%%%%%%%%%%%%%%%%%%%%%%%%%%%%%%%%%%%%%%%%%%%%%%%%%%%%%%%%%%%

\section{Simulating the Unruh effect with cold atoms}
\label{sec:proposal}

In this section we present our proposal to study the Unruh effect for
Dirac fermions in an optical lattice, in one and two spatial
dimensions. The crucial idea behind our proposal is that all
measurements made by an accelerated observer on the Minkowski vacuum
of the Dirac field can be simulated by {\em quenching} the Dirac Hamiltonian
from the one in  Minkowski spacetime to the one in Rindler spacetime, which amounts to quench
the tunneling amplitudes from constant to properly position-dependent values. 
As a by far non-trivial consequence of thermalization theorem ({\it cf.} Sect. \ref{sect:thermalization}), the Minkowski
vacuum will now be seen as a thermal state in Rindler, which we will
subsequently probe with a suitable analogue of De Witt detectors,
yielding the local fluctuation spectrum predicted by the Unruh effect.
As a thermal state corresponds only to populations of Rindler modes, the Minkowski vacuum is stationary, that is to say
that is invariant under time translations in Rindler spacetime (this property is not so surprising because 
Rindler time translations correspond to Lorentz boosts in the original Minkowski coordinates). 

We start by providing an overview of the experimental procedure. Our
scheme relies crucially on one-particle excitation spectroscopy, which
we discuss in detail. The robustness of our scheme is then validated
by performing a numerical simulation of the response function in
realistic experimental conditions. We conclude by proposing an
experimental implementation of the protocol which is accessible using
state-of-the-art techniques.

\subsection{Strategy}

As explained in Sect. \ref{sec:Unruh_effect}, the observation of the
Unruh effect requires a measurement of the Wightman two-point
correlation function in the frequency domain \eqref{detector.response} for an accelerated observer in the
Minkowski vacuum. In other terms, we have to measure the Fourier
transform of two-point correlations in time. For a Dirac system as
the one we consider, the Minkowski vacuum is the Fermi sea and
what needs to be measured is the overlap between the state
corresponding to a one-hole excitation at different times.
Furthermore, this one-hole excitation must follow an accelerated
trajectory.

Traditionally, an accelerated observer is considered in order to
observe the Unruh effect, with the
Minkowski vacuum at rest and the one-hole excitation moving.
% AC: comment to [31]
For instance, this is the approach considered in \cite{Retzker.PRL.08}, where
the one-hole excitation is created by the coupling to an impurity.
However, due to the equivalence principle, the measurement can actually be done in any
reference frame. We choose to perform it in the rest frame
of the observer and the one-hole excitation. There, the time
evolution is governed by the Dirac Hamiltonian in Rindler
spacetime (Eq. (\ref{rindler.ham})), and the response function is
simply the overlap between the one-hole excitation at rest at
different times. The measurement of the Wightman spectral function
can then be interpreted as the creation of a one-hole excitation
at a fixed location $x_0$ in the Fermi sea, the evolution of this
state with the Dirac Hamiltonian in Rindler spacetime for a time
$t$, and the creation of a particle at $x_0$. This is exactly what
one-particle excitation spectroscopy, a standard technique in cold atom experiments,
determines \cite{Dao.07}.

Thus, our protocol to observe the Unruh effect consists of three
steps:
\begin{enumerate}
\item \emph{Preparation of the Minkowski vacuum} by achieving the ground
state of the Dirac Hamiltonian with a uniform Fermi velocity
(Dirac Hamiltonian in Minkowski spacetime);
\item \emph{Quench to an accelerated frame} governed by
the Dirac Hamiltonian with a spatially dependent Fermi velocity
(Dirac Hamiltonian in Rindler spacetime). The quench introduces
an event horizon in the middle of the gas, effectively disconnecting it in two halves;
\item \emph{Measurement of the Wightman function} in the accelerated
(Rindler) frame using local one-particle excitation spectroscopy at point $x_0$.
\end{enumerate}

The first two steps provide a convenient method for preparing the
Minkowski vacuum as the ground state of a Hamiltonian which can be
easily implemented experimentally, and for making it evolve into an
accelerated (Rindler) frame. The third step, local one-particle
excitation spectroscopy, is the crucial ingredient of our proposal. It
creates a one-hole excitation in the gas, whose dynamics in the
accelerated (Rindler) frame produces the Bogoliubov transformation
\eqref{bogoliubov.transf}.  And it is this transformation which is
responsible for the thermalization theorem and the Unruh effect. Given
its importance, we describe it in detail in the next section.

\subsection{Measurement of the Wightman function} \label{sect:signature}

Our proposal for observing the Unruh effect relies on the use of
one-particle excitation spectroscopy for measuring the Wightman
function. This technique consists in transferring a fraction of atoms
of the gas to an auxiliary energy band which is initially unoccupied
and has a considerably smaller bandwidth, so that it can be
neglected. The process requires a field coupling both bands, and can
be implemented in a variety of fashions (radio-frequency, one-photon
or two-photon laser transitions) depending on the atomic species
chosen. In our case, we require the process to be local, since
the Wightman function is defined locally (at point $x_0$) and the
Unruh temperature varies as a function of the distance to the horizon.

If we consider the ensemble of the two bands as an effective two-level
system, the effect of the coupling can be modeled in the interaction
picture as

\begin{equation}
W_{x_0}(t) = W_0 (\e^{i \omega t} b_{x_0}^\dagger c_{x_0}(t) +
\e^{-i \omega t} c_{x_0}^\dagger(t) b_{x_0}),
\label{eq:raman_spectro}
\end{equation}
where $\omega$ represents the detuning between the frequency of the
field and the energy of the auxiliary band where the operator
$b_{x_0}^\dagger$ ($b_{x_0}$) creates (destroys) an atom, and we
assume an integration over all momenta. Since the measurement is performed after the quench, the
operator $c_{x_0}^\dagger(t)$ ($c_{x_0}(t)$) evolves with the Rindler
Hamiltonian.

Now, let us compute the occupation of the auxiliary band at a later
time.  As it is highly excited, we can assume it to be initially
empty. The initial state is thus

\begin{equation}
\ket|\Phi>_0=\ket|\Phi(t=0)> = \ket|0>_b\, \ket|\Omega> .
\end{equation}
Taking a sufficiently small coupling $W_0$ allows us to treat
\eqref{eq:raman_spectro} at first order in perturbation theory. We
find

\begin{align}
  \ket|\Phi(t)> &\sim \ket|0>_b\, \ket|\Omega> \\
  &+ W_0\int_0^t \d t'\; \e^{i \omega t'}\;
  b_{x_0}^\dagger \ket|0>_b\, c_{x_0}(t)
\ket|\Omega>.\nonumber
\end{align}
Then, the occupation $N_b$ of the auxiliary state for $t\gg 1/\omega$
is

\begin{align}
N_b &=\bra<\Phi(t)|b_{x_0}^\dagger b_{x_0} \ket |\Phi(t)> \nonumber\\
&= W_0^2 \int_0^t \int_0^t \d t' \d t'' \e^{i \omega (t'-t'')}
\bra<\Omega| c_{x_0}(t'') c_{x_0}(t')\ket|\Omega>\nonumber\\
&\propto \int_{-t}^t  \e^{i \omega t'}
\bra<\Omega| c_{x_0}(t') c_{x_0}(0) \ket|\Omega>\propto G(\omega),
\end{align} \label{eq:Nb}
where we have used translation invariance in time. The calculation
above not only demonstrates that local one-particle excitation
spectroscopy measures the Wightman spectral function. It also clearly
shows that it is the time evolution under the Rindler Hamiltonian
which is responsible for the observed thermal response.
We have calculated the occupation $N_b$ assuming that the pulse started at $t=0$, i.e., immediately after the quench --we have used the original Minkowski
vacuum to start with--. However, the integral \eqref{eq:Nb} depends only on the duration of the pulse, and not on the actual moment at which the pulse starts. 
This is an experimental manifestation of the stationary of Minkowski vacuum in Rindler spacetime and of the time invariance of the associated populations as Rindler
particles. 

\subsection{Validity range of the scheme} \label{sect:signature}

Let us now discuss in detail the range of validity of our approach and
review some possible limitations.

In our scheme we do not implement the Dirac Hamiltonian (both in
Minkowski and Rindler spacetimes) in the continuum, but only a lattice
version of it.  This introduces a characteristic length scale in the
system, the lattice spacing, and an associated UV energy cut-off.
Measurements of the Wightman function below this length scale are not
meaningful. However, the finite spatial resolution that one-particle
excitation spectroscopy will have in the experiment naturally smears
out these discretization artifacts. We will show below that a
measurement of the response function convolved over two lattice sites
is sufficient to suppress most of them.  Another consequence of
implementing the Dirac Hamiltonian in an optical lattice is that the
relativistic dispersion relation only holds in a certain range of
energies, in the vicinity of the Dirac points. Thus, the measurements
must be restricted to this energy range, which is given by the local
tunneling rate. This limitation is common to other proposals for
simulating relativistic effects with cold atoms. For instance, using
the Bogoliubov excitations of a Bose-Einstein condensate as relativistic
particles is only valid in the phonon-like regime of the Bogoliubov
dispersion relation, and breaks down away from it.

Our protocol relies on a change of reference frame, from a rest frame
to an accelerated one.  This step is done by quenching the Hamiltonian
from Minkowski to Rindler spacetime.  The change of reference frame
should be instantaneous, an approximation which is valid if the quench
time is much shorter than the smallest characteristic timescale of the
system given by the inverse of the largest tunneling rate. We will
see in Sect. \ref{sect:implementation} that experimentally this is a
reasonable assumption.  The main effect of the quench is to introduce
an artificial horizon in the middle of the lattice that effectively
disconnects the left and right halves. Placing the horizon exactly in
the middle of the system is important to minimize the distortions
induced by the finite system size. Finally, let us remark that
quenching the Hamiltonian of a quantum system normally triggers a
temporal evolution of its initial state. In our case, however, as we have observed at the end 
of Sect. \ref{sect:thermalization} and further argued in this section, the
Minkowski vacuum looks stationary to the accelerated observer and this
dynamics is absent. Indeed, the Rindler Hamiltonian is
proportional to the entanglement Hamiltonian of both halves of the
system \eqref{entanglement_Hamiltonian}. Thus, the density matrices of
both halves are time-invariant as they are diagonal and depend only on the populations.

The Unruh effect implies that measurements of the Wightman function at
different distances from the horizon, and thus different
accelerations, will yield different values of the Unruh
temperature. In order to compare these measurements, the rates should
be measured with respect to the proper time $\tau$. For the Rindler
metric, Eq. \eqref{optical.metric}, $\tau = \xi t$ with $\xi \propto
|x|$. Thus, frequencies must be scaled by $1/\xi$. At the same time,
the Fourier transformation of the Wightman function $G(t)$ has to be
performed with respect to the proper time.  In the frequency domain it
is then given by $\xi G(\omega/\xi)$.

Finally, up to now we have been assuming that the Minkowski vacuum can
be exactly realized in the experiment. Or, in other terms, that it is
possible to prepare perfectly the ground state of the Dirac
Hamiltonian in the homogeneous tunneling lattice at half filling
(exactly up to the Dirac points).  In real experiments, however, the
actual temperature of the fermionic gas will not be zero but rather on
the order of the tunneling rate.  We will show that the signatures of
the Unruh effect can still be appreciated when starting with a
finite-temperature sample.

\subsection{Numerical simulations}

In order to validate our scheme and address the effects presented
above we have performed numerical simulations, which we present in
this section.

The calculations have been done using the $\pi$-flux realizations of
the Dirac Hamiltonian (Eqs. (\ref{pi.flux.Hamiltonian}) and
(\ref{Rindler.pi.flux.Hamiltonian})) in one and two spatial
dimensions.  In both cases we have simulated numerically the complete
scheme, starting with an initial state in Minkowski spacetime,
assuming an instantaneous quench, and computing then the Wightman
response function as will measured by one-particle excitation
spectroscopy (see Eq. \eqref{occupation.thermal}).

In the calculations we fix the system sizes $L_x$, $L_y$. The natural energy scale of
the system is the bandwidth of the Dirac Hamiltonian in Minkowski
spacetime (proportional to the tunneling strength $t_0$ in
\eqref{pi.flux.Hamiltonian}). Therefore, all energies
(i.e. frequencies and temperatures) are measured in units of
$t_0$. The amplitude $t'_0$ that characterizes the tunneling strength
of the Dirac Hamiltonian in Rindler spacetime
\eqref{Rindler.pi.flux.Hamiltonian} is in principle arbitrary due to
the overall scale invariance of the Rindler space. We choose it so
that the maximal tunneling rate is equal to the Minkowski value, $t'_0
= 2 t_0/L_x$. The lattice spacing is fixed as $d=1$ and we attach
$(x,y)=(m,n)$ coordinates to each site in a symmetric way with respect
to the horizon, i.e.  the $x=0$ line. Thus, while $n$ always runs over
integers, $n=1,2\dots,L_y$, $m$ runs over integers for $L_x= 2N+1$
odd, $m= -N,-N+1\dots,N$, and over half-integers for $L_x=2 N$ even,
$m= -N+1/2,-N+3/2\dots,N-1/2$. Note that fixing the horizon exactly in
the middle of the gas is important to minimize the distortions
introduced by the finite size of the system.

As we mentioned in the previous section, in a discrete realization of
the Dirac Hamiltonian only measurements performed at length scales
above the lattice spacing $d$ are meaningful.  The finite spatial
resolution of the measurements will automatically perform the required
coarse-graining. We simulate it numerically by considering a
convolution of the Wightman function \eqref{detector.response.full}
with a Gaussian of standard deviation corresponding to $2$ lattice
sites along the $x$ direction. The raw data obtained before the
convolution, and further details concerning it are included in
Appendix \ref{sect:A3}.

\begin{figure}[h!]
%\rput(0,5.5){(a)}
\includegraphics[width=8.5cm]{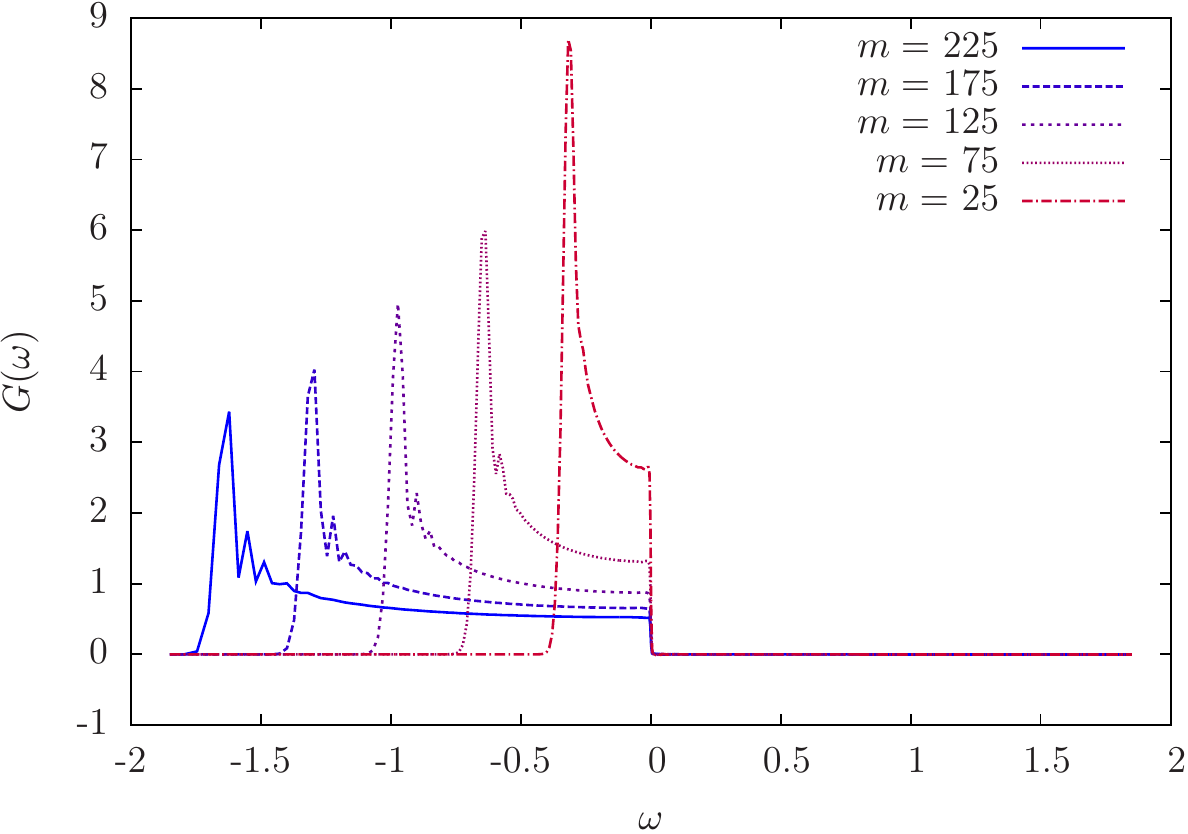}
%\rput(0,5.5){(b)}
\includegraphics[width=8.5cm]{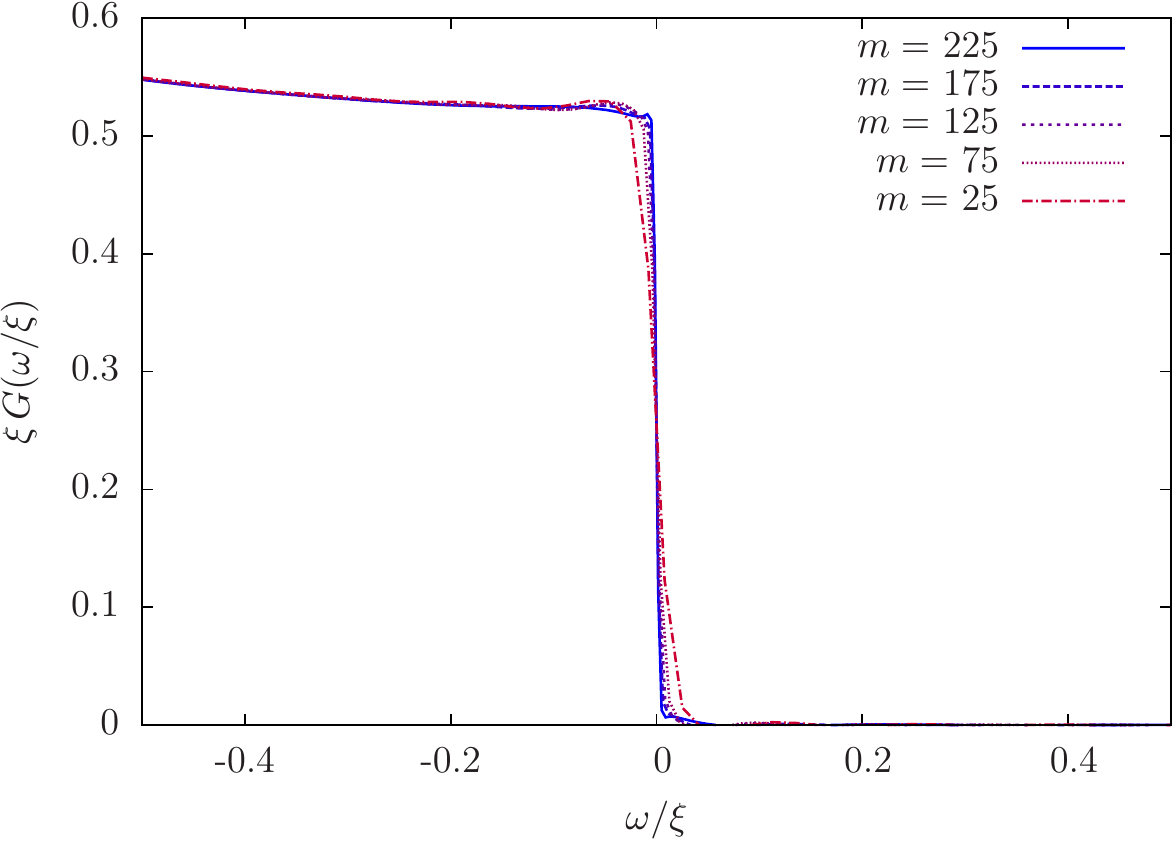}
\caption{(a) Wightman response function in the frequency domain for a
  1D system of size $L_x=500$ after the quench. The colors denote
  different distances to the horizon, $m$, expressed in lattice sites:
  blue (lower) is far away and red (taller) is closest to it.  (b)
  Wightman response function of the same system in the frequency
  domain, measured with respect to the proper time $\tau= \xi t$. The
  proper frequency is $\omega/\xi$, while taking the Fourier transform
  with respect to $\tau$ requires rescaling $G(\omega) \to \xi
  G(\omega/\xi)$. As explained in the main text, $\xi G(\omega/\xi)$
  represents what a static De Witt detector in Rindler spacetime would
  observe.  Notice that the curves collapse to Fermi-Dirac
  distributions of increasing temperatures as we approach the
  horizon. For $|\omega/\xi|>1$ lattice artifacts (deviations from the
  relativistic dispersion relation) distort the response.}
\label{fig:unruh_1d}
\end{figure}

The frequency dependence of the response is evaluated at five
different positions, at linearly increasing distances from the
horizon. The top panel of Fig. \ref{fig:unruh_1d} (a) shows the
convoluted results obtained for a 1D system of size $L_x=500$. The
tallest (red) curve is the closest to the horizon, and the lowest
(blue) one is the most distant. For frequencies close to zero
(i.e. the Fermi energy) the response functions all have a behavior
resembling a Fermi-Dirac distribution, with strong lattice artifacts
at large negative frequencies.

In order to compare the different results, we rescale the curves with
respect to the proper time $\tau$.
The proper frequency is then $\omega/\xi$, and the proper rate of
detection is $\xi G(\omega/\xi)$. For our choice of units
$t_0=d=1= \tfrac {L_x}2 t'_0$, $\xi = 2 x/L_x$. Thus, $\xi=0$
corresponds to the horizon and $\xi=1$ to the edge of the system.
Fig. \ref{fig:unruh_1d} (b) presents the same curves as Fig. \ref{fig:unruh_1d} (a),
but in rescaled units. For frequencies close to $\omega=0$, they
reproduce Fermi-Dirac distributions whose temperature increases as we
approach the horizon. The distributions are
{\em not} normalized, since they are defined up to a global
constant.

Notice that in Fig. \ref{fig:unruh_1d} we have restricted the
displayed frequency range to the regime where the energies are lower
than the local tunneling range $|\omega|<|t(x)|$, since it is only
there that the dispersion relation remains linear and the description
of the particles in terms of Dirac fermions is valid. In rescaled
units this condition becomes $|\omega/\xi|<1$.  In the following we
will restrict ourselves to this frequency range.

\begin{figure}[h!]
%\rput(0,5.5){(a)}
\includegraphics[width=8cm]{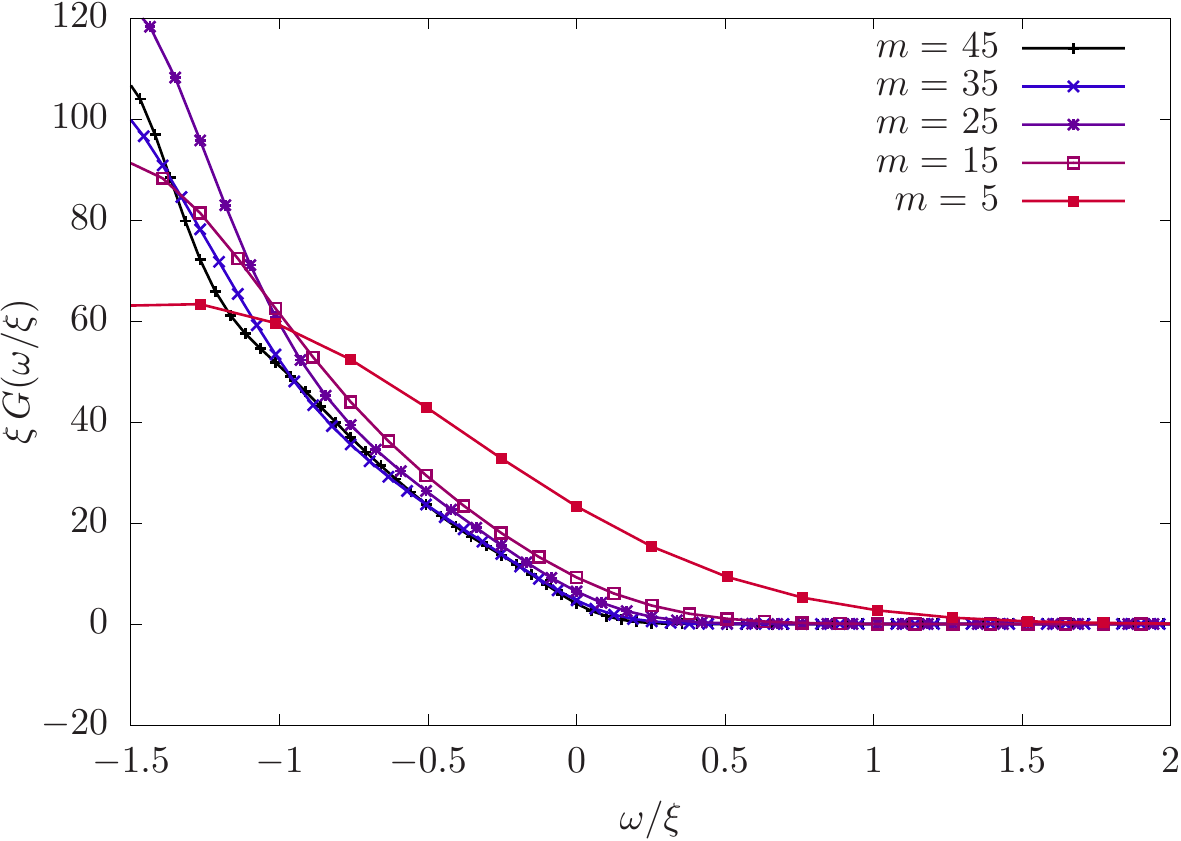}
%\rput(0,5.5){(b)}
\includegraphics[width=8cm]{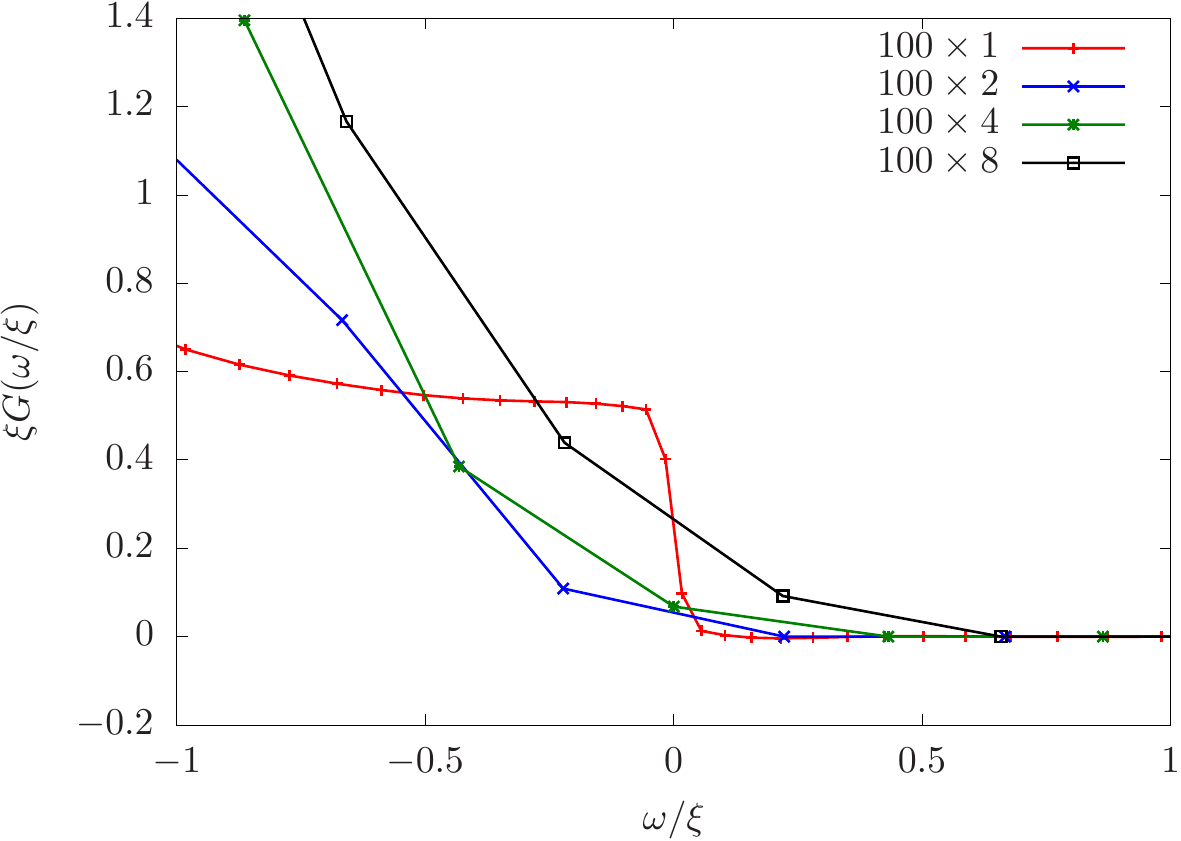}
\caption{(a) Wightman response function in the proper frequency domain
  for a 2D system of size $100\times 100$ after the quench. As
  previously, the colors denote different distances from the horizon,
  $m$, expressed in lattice units: blue (+ signs) is far away, red
  (full squares) is closest to it. (b) Same for strips of different
  widths, $100\times 1$ to $100\times 8$.}
\label{fig:unruh_2d}
\end{figure}

Fig. \ref{fig:unruh_2d} shows the corresponding results for a 2D
system, which differs strongly from its 1D counterpart as predicted by
Takagi's inversion theorem. Fig. \ref{fig:unruh_2d} (a) shows the
rescaled response function for a $100\times 100$ lattice, measuring at
linearly increasing positions from the horizon as in the 1D case. The
displayed curves include the spatial Gaussian convolution, along with
an energy coarse-graining $\Delta\omega=0.2$.  The latter simulates
the finite energy resolution of the measurement, limited by the finite
system size. The raw data, prior to convolution and coarse-graining is
presented in Appendix \ref{sect:A3}.  As predicted by Takagi, the
results are now similar to a Bose-Einstein distribution.

Finally, in Fig. \ref{fig:unruh_2d} (b) we study the transition between 1D and
2D, by showing the rescaled response functions for a set of
lattices with dimensions $100\times 1$ (red), $100\times 2$,
$100\times 4$ and $100\times 8$ (black), always measured at a
point $25$ lattices sites away from the horizon. Notice that the
Fermi-Dirac distribution disappears very fast when we increase the
transverse dimension $L_y$.

\begin{figure}[h!]
%\rput(0,5.5){(a)}
\includegraphics[width=8cm]{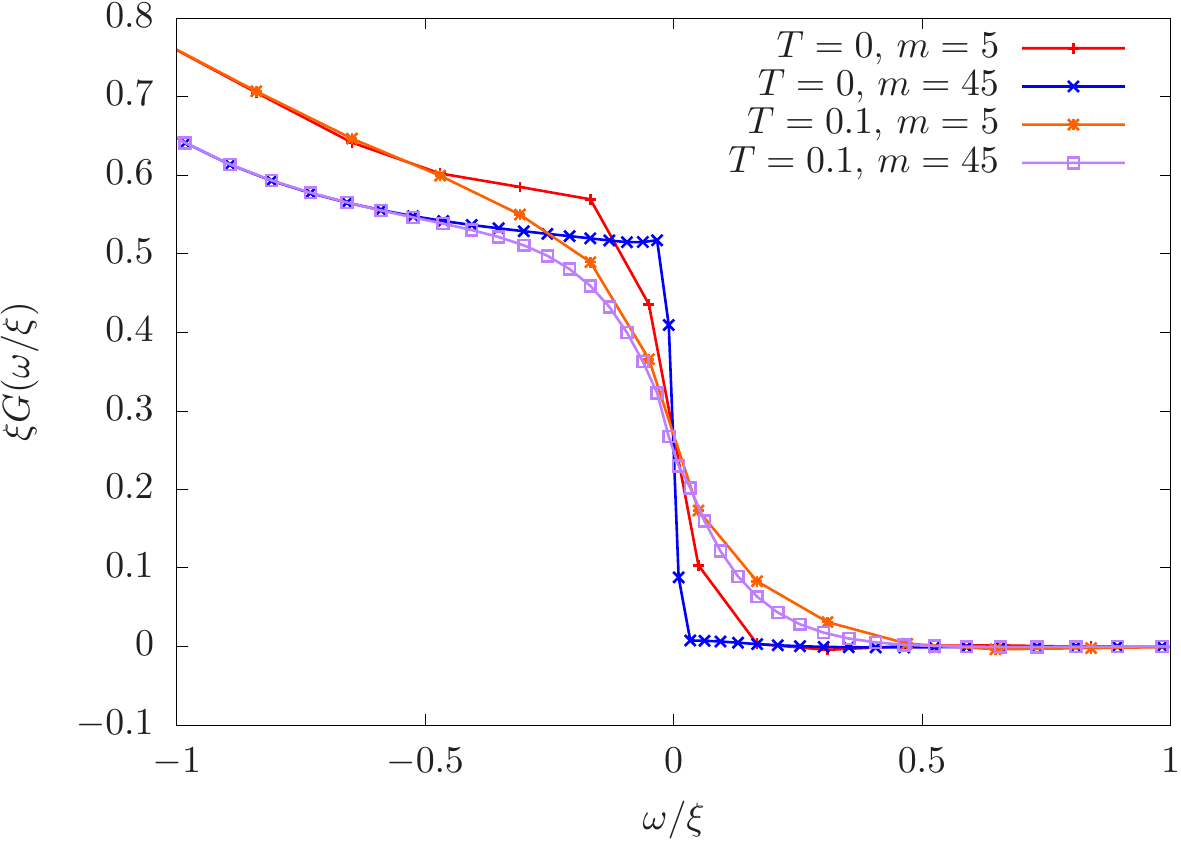}
%\rput(0,5.5){(b)}
\includegraphics[width=8cm]{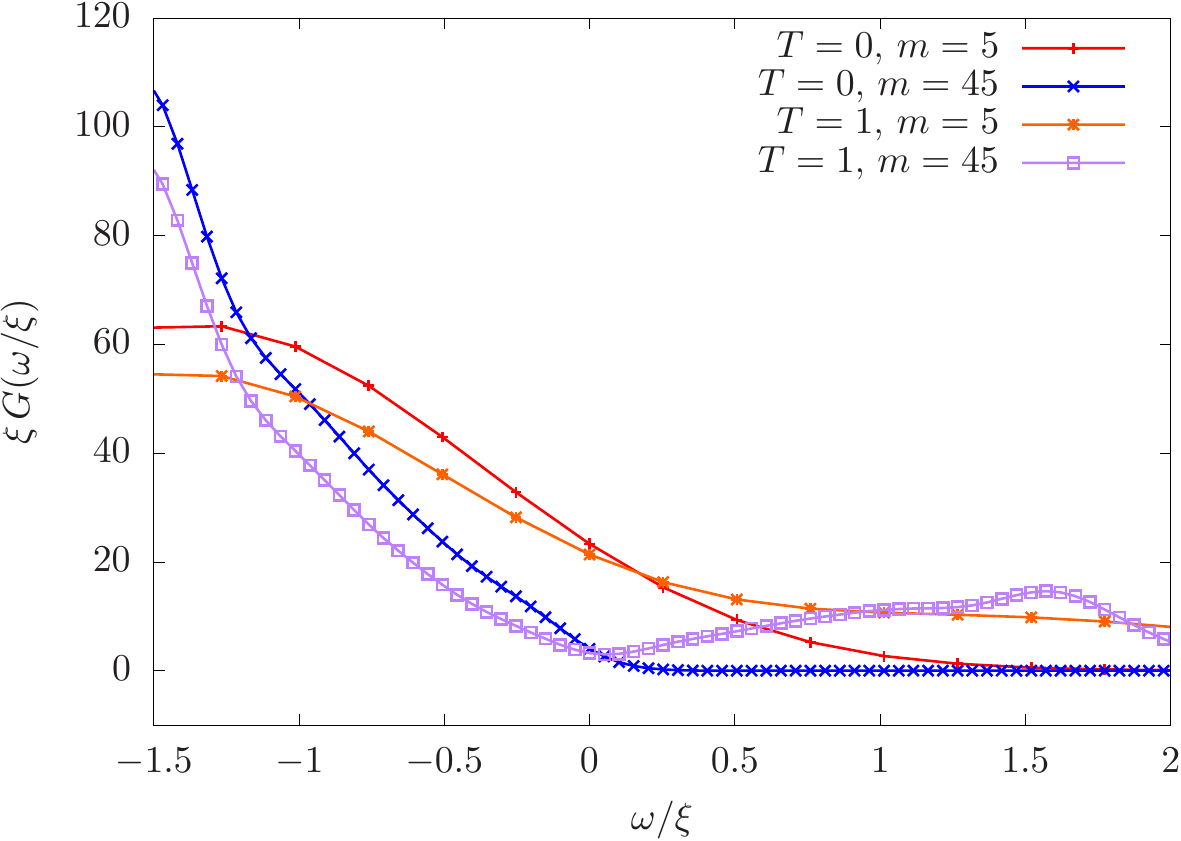}
\caption{(a) Comparison between the Wightman response functions in the
  proper frequency domain for a 1D system of size $L_x=100$ after the
  quench, for two different physical temperatures, $T=0$ and $T=1/10$,
  and at two different locations, near and far from the horizon.  The
  two response functions at $T=1/10$ resemble Fermi-Dirac
  distributions at finite temperature, but the one measured closer to
  the horizon corresponds to a higher temperature than the one
  measured far away. (b) Same comparison for a 2D system of size
  $100\times 100$. The response curves keep their global features when
  increasing the system temperature from $T=0$ to $T=1$.  The response
  function far from the horizon, $m=45$, shows a local maximum at
  positive frequencies before decaying. This behavior is analogous to
  the one expected for a thermal gas of Dirac fermions in the
  homogeneous tunnelling lattice (see the Appendix \ref{sect:A3},
  Fig. \ref{fig:2d_mink_gas}).  Indeed, in Rindler spacetime the limit
  $m\to \infty$ corresponds to zero acceleration and converges to the
  results of Minkowski spacetime.}
\label{fig:unruh_thermal}
\end{figure}

As a last step, we study the robustness of our protocol under
an increase in the physical temperature of the gas. This results in an imperfect
preparation of the Minkowski vacuum, which is the starting point of the protocol.
Fig. \ref{fig:unruh_thermal} (a) compares the rescaled response
functions for a 1D system at physical temperature $T=0$ and $T=1/10$,
measured at two different points, one close to the horizon and one far
from it. At $T=1/10$ the distributions are rounded near $\omega=0$,
but we can still see that the one closest to the horizon is more
curved and presents a larger probability for positive energy
excitations. The 2D case is more robust, as shown in
Fig. \ref{fig:unruh_thermal} (b). There, we can see that the rescaled
distributions at $T=1$ measured near and far from the horizon are
clearly distinguished, and keep the same global features than at
$T=0$. The explicit expression used to calculate the response functions for a
thermal gas is given in the Appendix \ref{sect:A3}.

% \FloatBarrier
%%%%%%%%%%%%%%%%%%%%%%%%%%%%%%%%%%%%%%%%%%%%%%%%%%%%%%%%%%%%%%%%%%%%%

\begin{figure*}
\includegraphics[width=1.99 \columnwidth]{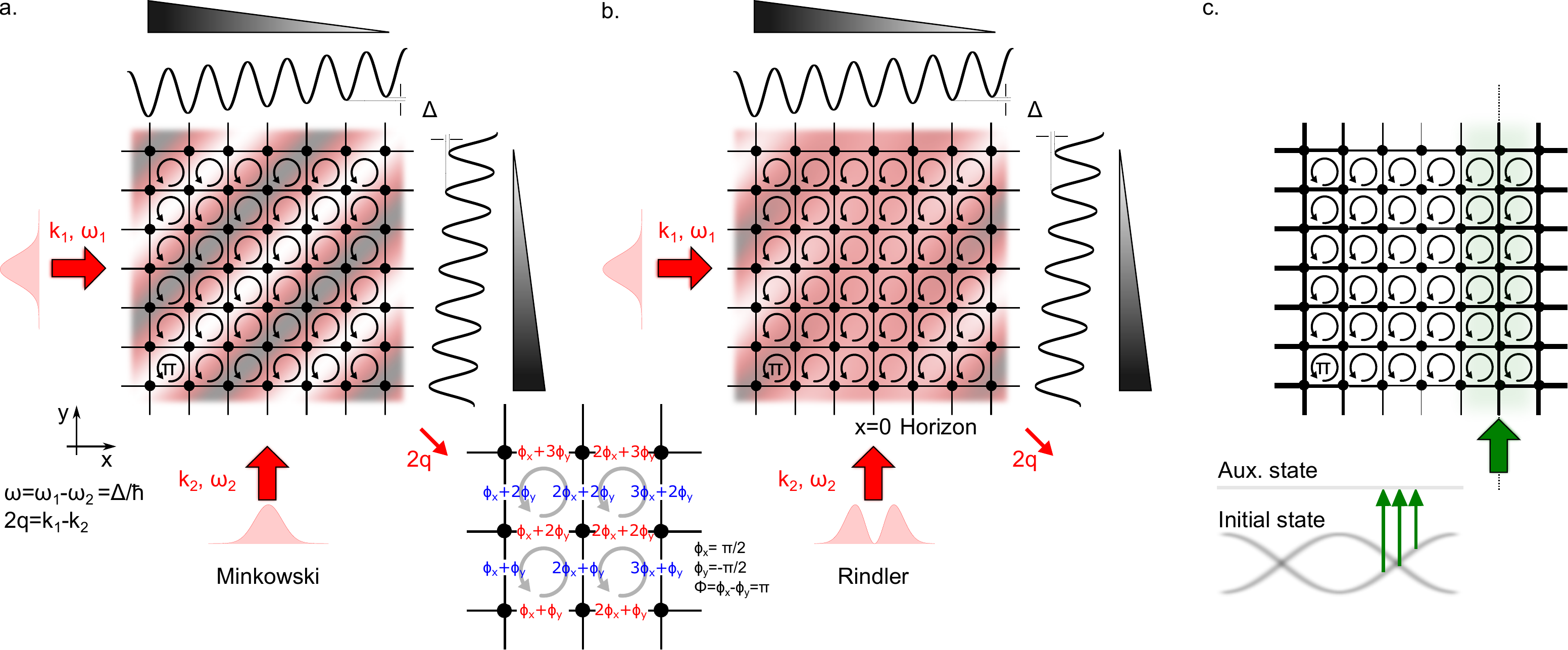}
\caption{\label{fig.implementationH} (a) Experimental scheme for
implementing the $\pi$-flux model in the symmetric gauge with
homogeneous tunneling amplitudes (Minkowski). A linear potential
gradient of amplitude $\Delta$ is superimposed along the diagonal
direction of a 2D lattice. Tunneling is restored using a pair of
Raman beams of frequencies $\omega_{1,2}$ and wavevectors
$\mathbf{k_{1,2}}$ which create a modulated potential of frequency
$\omega=\Delta/\hbar$ and wavevector $\mathbf{q}$ (red snapshot).
This leads to complex tunneling with the required spatial
dependence of the phase (inset). The tunneling amplitude is
homogeneous over the system when using Gaussian Raman beams of
large waist compared to the size of the cloud. (b) The Dirac
Hamiltonian in Rindler spacetime is realized when one of the Raman
beams has instead a TEM$_{10}$ Hermite-Gauss spatial mode, leading
to a linear dependence of the tunneling amplitude with respect to
$x=0$ (event horizon). (c) The measurement of the detector
response function could be realized by local band spectroscopy,
using a spectroscopy beam focused at different distances to the horizon (green).}
\end{figure*}

\subsection{Experimental implementation} \label{sect:implementation}

Our proposal to implement experimentally the Dirac Hamiltonian in
Minkowski and Rindler spacetimes is based on the recent
experimental realizations of the Hofstadter model with ultracold
atoms \cite{Aidelsburger.13, Miyake.13, Kennedy.15}, but in the
symmetric gauge and using fermionic atoms instead.

As sketched in Fig. \ref{fig.implementationH}, a two-dimensional
square lattice with bare tunneling matrix elements $J$ along the
$x$ and $y$ directions, and lattice spacing
$d=\lambda_{\mathrm{L}}/2$ (where $\lambda_{\mathrm{L}}$ is the
wavelength of the lattice beams), is subjected to a potential
gradient oriented along the diagonal direction of the lattice.
This leads to an energy offset between neighboring sites
$\Delta\gg J$ which inhibits tunneling. The offset value could
depend on the state of the atom, but should be identical along the
$x$ and $y$ directions. A pair of Raman laser beams collinear with
the lattice beams, of wave vectors $\mathbf{k_{1,2}}$ and
frequencies $\omega_{1,2}$ result in an additional optical
potential
\begin{equation}
V_K(\mathbf{r})\propto
\frac{V_K^0(\mathbf{r})}{2}\cos(\mathbf{q}\cdot\mathbf{r}+\omega
t),\label{eq:rampot}
\end{equation}
with $\mathbf{q}=\mathbf{k_1-k_2}$ and $\omega=\omega_1-\omega_2$.
The potential amplitude $V_K^0({\bf r})$ is assumed to be a slowly
varying function of $\mathbf{r}$. The effect of the Raman beams is
to restore tunneling along the two directions when the condition
$\omega=\Delta/\hbar$ is fulfilled, but with a spatial dependence
of the phase. In the high frequency limit $\hbar\omega\gg J$, the
system is then described by the effective Hamiltonian
\begin{multline}
H=-\sum_{m,n} \left( t(m+\tfrac 12,n)\,
\mathrm{e}^{i\phi_{m,n}}c^{\dagger}_{m+1,n}c_{m,n}\right.\\\
\left.+t(m,n)\, \mathrm{e}^{i\phi_{m,n}}
c^{\dagger}_{m,n+1}c_{m,n}\right)+\mathrm{H.c.}\,. \label{implementedHam}
\end{multline}

Here the phase factor is
$\phi_{m,n}=\mathbf{q}\cdot\mathbf{r}=m\phi_x+n\phi_y$.
The Dirac Hamiltonians \eqref{pi.flux.Hamiltonian} and \eqref{Rindler.pi.flux.Hamiltonian} are special cases of \eqref{implementedHam}.
For the
Raman laser propagation directions displayed in Fig.
\ref{fig.implementationH} and a Raman laser wavelength
$\lambda_{\mathrm{R}}=2\lambda_{\mathrm{L}}$, the phases are
$\phi_x=-\phi_y=\pi/2$, which corresponds to the $\pi$-flux
Hamiltonian in the symmetric gauge. The laser assisted tunneling
amplitudes are then given by
\begin{multline}
t(m,n) \simeq t \mathcal{J}_1(V_K^0(m\,  d, n\,  d)/\sqrt{2}\Delta)\\
   \simeq t V_K^0(m\,  d, n\, d)/2\sqrt{2}\Delta, \label{eq:ramtun}
\end{multline}
where $\mathcal{J}_1(x)$ is the Bessel function of the first kind.
This expression is valid in the limit $\Delta\gg V_K^0(m\,  d, n\,
d)$ and for slowly varying $V_K^0(\mathbf{r})$, which allows to
use as average amplitude of the potential its value at the center
of the link. This scheme allows for the simulation of the whole
family of optical metrics \cite{Cvetic.12} considered by some of
us in \cite{Boada.11}, and also of extensions of this family to
include a mild time dependence in the metric \cite{Minar.15}.

The realization of the Dirac Hamiltonian in Minkowski spacetime
requires laser-assisted tunneling amplitudes $t(m,n)=t_0$
homogeneous across the cloud, which corresponds to a constant
value of the Raman optical potential amplitude. This could be
realized using Gaussian Raman beams of waist $w_0$ much larger
than the cloud size (see Fig. \ref{fig.implementationH}a). In
order to implement the Dirac Hamiltonian in Rindler spacetime, we
need instead tunneling amplitudes which increase linearly along
the $x$ direction, $t(m,n)=t'_0 m$ or, equivalently, a Raman
optical potential amplitude proportional to $x$. Using a
TEM$_{10}$ Hermite-Gauss mode \cite{footnote1} for the $y$ Raman
beam results in the large beam limit in a Raman optical potential

\begin{equation}
V_K^R(\mathbf{r})\propto \sqrt{2}\left(\frac{x}{w_0}\right)
V^0_{K}\cos(\mathbf{q}\cdot \mathbf{r}+\omega t),
\end{equation}
which, as follows from \eqref{eq:ramtun}, leads to the required
spatial dependence of $t$ (see Fig. \ref{fig.implementationH}b).
% Leticia's answer to the first referee: how we obtain a TEM01. In footnote because it is very technical and quite obvious.
% Timescale remark is more pertinent and goes in main text.
The quench between the two situations (Minkowski and Rindler)
could be performed by a sudden change of the mode of the $y$ Raman
beam, on a timescale of $\sim 10\,\mu$s. This is well below the
shortest timescale of the system, given by the inverse of the
highest tunneling rate, which will typically be on the order of
$\sim 10$ ms. We thus consider the quench as instantaneous.
Finally, this scheme can be easily modified, adding for example a
superlattice potential along the $y$ direction, in order to
interpolate between the 1D and 2D situations and observe the
inversion of statistics.

For measuring the Wightman function $G(\omega)$ we propose to
perform local spectroscopy of the energy bands
and determine their occupation as a function of energy. This
information is contained in the transfer rate from an atomic state
experiencing the Dirac Hamiltonian in Rindler spacetime after the
quench, to an auxiliary atomic state with a different dispersion
relation. It could thus be measured using one-particle excitation
spectroscopy, as demonstrated in \cite{Stewart.08} and more
recently used to characterize spin-orbit coupled Fermi gases
\cite{Wang.12, Cheuk.12}. In order to perform local measurements
and determine the dependence of the detector response function
with the distance to the horizon, the transfer could be performed
using a spectroscopy beam \cite{Dao.07} focused at different $x$ positions
(see Fig. \ref{fig.implementationH}c). Note that the finite waist
of this measurement beam, larger than the lattice spacing, would
remove from the measurement some of the discretization artifacts
discussed previously, and is equivalent to the convolution
procedure used in the numerics (see Appendix \ref{sect:A3}). % Leticia: cite New Appendix A3 on convolution
Experimentally, the most challenging requirement for this local
spectroscopy scheme is to realize $\pi$-flux model using an atomic
species where one atomic state is subjected to the Dirac
Hamiltonian in Minkowski and Rindler spacetimes, whereas the
second (auxiliary) state experiences a different dispersion
relation. This situation could be achieved
exploiting the ground ($^1S_0$) and long-lived metastable excited state ($^3P_0$) of the
fermionic isotopes of Yb or Sr, since the different polarizability
of the two states leads to different optical potentials for a
broad range of lattice and Raman beam wavelengths $\lambda_L$ and
$\lambda_R= \tfrac{2\pi}{{\bf k}_{1,2}}$. $\lambda_L$ would then be chosen such that the auxiliary
band has a negligible bandwidth compared to the initial one. In this
scheme the potential gradient leading to the site offset $\Delta$
should be realized optically as well. The spectroscopy would be performed
using a single laser tuned to the clock transition. This ensures an
excellent energy resolution, below the tunneling energy scale, as
recently demonstrated experimentally in refs. \cite{Kolkowitz.16,
Livi.16}.
% Leticia's addition to answer referee 1

%%%%%%%%%%%%%%%%%%%%%%%%%%%%%%%%%%%%%%%%%%%%%%%%%%%%%%%%%%%%%%%%%%%%%%%

\section{Conclusions and further work}
\label{sec:conclusions}

We have developed a proposal for a quantum simulator of the Unruh
effect in 1D and 2D massless fermionic fields using ultracold
atoms in an optical lattice. The addition of interacting fields
and disorder is possible in our approach, which therefore
constitutes a full framework for the study of the theoretical
implications of quantum field theory in curved spacetime.
Moreover, our simulator provides a setting for the study of
relativistic quantum information theory in an experimentally
accessible system.

The implementation of this quantum simulator is within
experimental reach using state--of--the--art experimental
techniques. The detection methods proposed here are potentially
relevant also for detecting topological properties in simulators
of topological insulators and to assess the properties of quantum
systems out of equilibrium.

In this work we have restricted ourselves to the study of the
Rindler metric, i.e. Minkowski spacetime viewed by an accelerated
observer. Nonetheless, the formalism and experimental tools
described here may be extended to the study of more complex
spacetimes, for instance, non static or even non-stationary ones. Our
work can also be considered as the first mandatory step prior to the
inclusion of matter back-reaction in the artificial metric, and
to the simulation of dynamical gravity fields.
%%% Added sentence to clarify why such quantum simulator would be
%%% useful
Thus, the present work paves the way to experiments that are not only
fascinating \emph{per se}, but are also able to access phenomena that
are not fully understood theoretically, such as gravitating quantum
matter in interaction.

Finally, the recent conceptual developments towards a combination
of quantum mechanics and general relativity, such as quantum
graphity \cite{Hamma.PRD.10,Caravelli.12} or the Maldacena-Susskind notion of
relating entanglement and spacetime in order to avoid the firewall
problem \cite{Maldacena.13}, might also be amenable to quantum
simulation using a similar approach.

%%%%%%%%%%%%%%%%%%%%%%%%%%%%%%%%%%%%%%%%%%%%%%%%%%%%%%%%%%%%%%%%%%%

\begin{acknowledgments}
This work has been supported by Spanish MINECO (SEVERO OCHOA Grant
SEV-2015-0522, FOQUS FIS2013-46768, FIS2014-59546-P and
FIS2012-33642), the Generalitat de Catalunya (SGR 874),
Fundaci\'o Privada Cellex, DFG (FOR2414) and EU grants EQuaM (FP7/2007-2013 Grant
No. 323714), OSYRIS (ERC-2013-AdG Grant No. 339106), SIQS
(FP7-ICT-2011-9 No.  600645), QUIC (H2020-FETPROACT-2014 No. 641122)
and PCIG13-GA-2013-631633. The authors want to acknowledge A. Enciso,
I. Fuentes, J. Korbicz, J. Le\'on, D. Peralta, C. Sab\'{\i}n, and
G. Sierra for very fruitful discussions.
\end{acknowledgments}

\appendix

\section{Eigenstates of the Rindler Hamiltonian in 1D} \label{sec:1drindler}

Let us restrict ourselves to 1D. Since the $x<0$ and $x>0$ regions are
effectively separated, we may restrict ourselves to the right
half-line. Consider the spinless 1D version of
Eq. \eqref{diracEq.xpx}, $H_{R(1D)}=\sqrt{x}p\sqrt{x}$. Notice that
$x\pl_x=\pl_{\log(x)}$. Let us define $u\equiv\log(x)$, taking the
horizon to $-\infty$. So, $H_{R(1D)}=-i(\pl_u + 1/2)$. The eigenvalue
equation is

\begin{equation}
-i\(x\pl_x + 1/2\)\psi(x)=-i\(\pl_u + 1/2 \)\psi(u)= \omega\psi(u).
\label{xp.ham}
\end{equation}

\begin{figure}
\includegraphics[width=8cm]{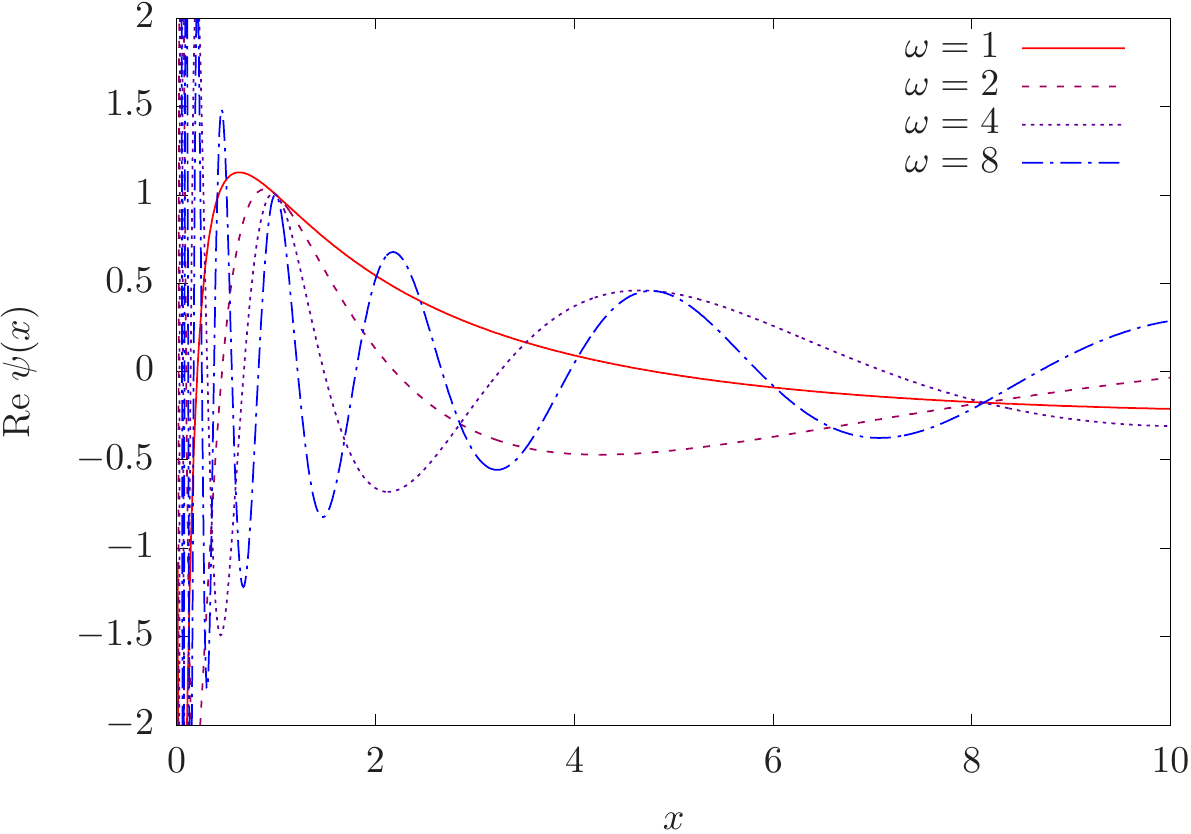}
\caption{One-dimensional Rindler modes, Eq. \eqref{rindler.sol}, for
  different values of $\omega$.}
\label{fig:solrindler}
\end{figure}

The solutions to that equation have the form

\begin{equation}
\psi(u)=A\; \exp \left[\(i\omega-\frac 12 \)u\right]=A\;x^{i\omega-1/2},
\label{rindler.sol}
\end{equation}
so they are plane waves in $u=\log(x)$. Fig. \ref{fig:solrindler}
shows the behavior of these wavefunctions.

In order to ensure that the Hamiltonian is truly Hermitian, we can
check that the eigenfunctions corresponding to different
eigenvalues are orthogonal. Indeed, they are:

\begin{eqnarray}
&\int_0^\infty \d x \exp\( (-i\omega-1/2)u \) \exp\( (i\omega'-1/2) u\)=
\nonumber \\
&\int_{-\infty}^\infty \d u \exp(-i\omega u) \exp(i\omega'u)
=\delta(\omega-\omega').
\end{eqnarray}

Let us insert the spinor structure. For $x>0$,

\def\spinor#1#2{\left(\begin{matrix}#1\\#2\\\end{matrix}\right)}

\begin{equation}
-i\( x\pl_x + 1/2 \)\sigma_x \spinor{\psi_1}{\psi_2}= \omega
\spinor{\psi_1}{\psi_2},
\end{equation}
which leads to $(\pl_u+1/2)^2\psi_1=-\omega^2\psi_1$, and an
equivalent equation for $\psi_2$. The solution is very similar to
the non-spinorial case

\begin{eqnarray}
\spinor{\psi_1(x,t)}{\psi_2(x,t)} = A \spinor{1}{\pm 1}
x^{i\omega-1/2} e^{-i\omega t}.
\label{sol.xpx.1d.x}
\end{eqnarray}

\subsection{Discretization of the Rindler Hamiltonian}
\label{subsec:discr}

The implementation of an analogue of equations
\eqref{dirac.Eq.mink} or \eqref{diracEq.xpx} in an optical lattice
requires a suitable discretization. In this section we will
discuss the 1D case.

Let us discuss how to discretize $H_{R(1D)}=\sqrt{x} p \sqrt{x}$,
the 1D Rindler Hamiltonian, appropriately. Consider an open 1D
lattice with spacing $d$, and lattice points $x_m= m d
$, with $m\in\{-(L-1)/2,\cdots,(L-1)/2\}$ and even $L$. Thus, the
wavefunctions only take components $\psi_m\equiv \psi(x_m)$. Let
us use a central differences discretization for $p=-i\pl_x$, i.e.,
$(p\psi)_m=-i\(\psi_{m+1}-\psi_{m-1}\)/(2d)$. Let us call
$R$ the discrete version of the $H_{R(1D)}$, for later
convenience:

\begin{align}
(R\psi)_m =& \sqrt{x_m} (p\sqrt{x}\psi)_m \cr
=&- \frac i{2}
\(\sqrt{m (m+1)}\psi_{m+1} - \sqrt{m (m-1)}\psi_{m-1} \) \cr
=& \sum_{m'} R_{m,m'} \psi_{m'} .
\label{discretizing}
\end{align}

Thus, the matrix entries for the Hamiltonian $R_{m,m'}$ are
non-zero only when the difference between the spatial indices is one:
$|m-m'|=1$. In that case,

\begin{equation}
R_{m,m+1}=-\frac{i}{2}\; \sqrt{m (m+1)}.
\label{matrix.entries}
\end{equation}
This means that the tunneling between sites $m$ and $m+1$ must be
$-(i/2)\;\sqrt{m(m+1)}$, independently of $d$. This is not surprising,
since both $R$ and Rindler spacetime are {\em scale invariant}. A good
approximation is obtained by replacing the {\em geometric mean} by the
{\em arithmetic mean}: $R_{m,m+1} \approx -(i/2)\;(m+1/2)$.

The discrete Hamiltonian \eqref{matrix.entries} can be
analytically diagonalized \cite{Sierra.05}. Its spectrum becomes
continuous with constant energy level density as $L\to\infty$, but
the convergence rate is very slow: the distance between levels
scales as $\approx 1/\log(L)$.

Let us now focus on the 2D case. In order to formulate the Rindler Hamiltonian
on a square lattice it is convenient to start with the symmetric continuous formulation
\eqref{symHRcont}. Explicitly, by writing the spinor in terms of its chiral components,
$\psi(x,y)=\left(\begin{smallmatrix} a(x,y)\\ b(x,y)
                     \end{smallmatrix}\right)$, we have
\begin{multline}
H_R= \frac i2 \int \d x\d y \, |x|\left(\partial_x a^\dag(x,y) \right.\\
     \left. -i \partial_y a^\dag(x,y)\right) b(x,y) + {\rm H.c.}\,.
\end{multline}
We can now exploit the bipartition of the lattice for discretizing
separately the two chiralities in the two checkerboard sublattices
and write the kinetic term in terms of the tunneling between the
two
\begin{multline}
H_R= \frac i4 t'_0 \sum_{k,l}\left( |k+l+\tfrac 12|\, a^\dag_{k+l +1,k-l}+ \right.\\
                      - |k+l-\tfrac 12|\, a^\dag_{k+l - 1 ,k-l} - i \,|k+l|\, a^\dag_{k+l,k-l + 1 } \\
     \left. +i \,|k+l|\, a^\dag_{k+l,k-l - 1 } \right) b_{k+l,k-l} + {\rm H.c.}\, .
\end{multline}
At this point we notice that the above Hamiltonian can be
rewritten as the $\pi$-flux Hamiltonian once we do not distinguish
fermions in the different sublattices. Denoting the annihilation
(creation) operators by $c_{m,n}$ ($c^\dag_{m,n}$),
\begin{multline}
H_R= \frac {t'_0}2\sum_{m,n}\left( i\, |m+\tfrac 12|\, c^\dag_{m +1,n} + \right.\\
     \left.(-1)^{m+n}\, |m| \, c^\dag_{m, n + 1 }\right) c_{m,n} + {\rm H.c.}\, .
\end{multline}
By applying the gauge transformation
\begin{equation}
c^\dag_{m,n} \to e^{i\frac {\pi}4 \left(m^2-4n^2- 2 m n - 3 m +4 n\right)} c^\dag_{m,n},
\end{equation}
we can recover the $\pi$-flux Hamiltonian in the symmetric gauge
\eqref{Rindler.pi.flux.Hamiltonian}.

\begin{figure}[h!]
%\rput(0,5.5){(a)}
\includegraphics[width=8.5cm]{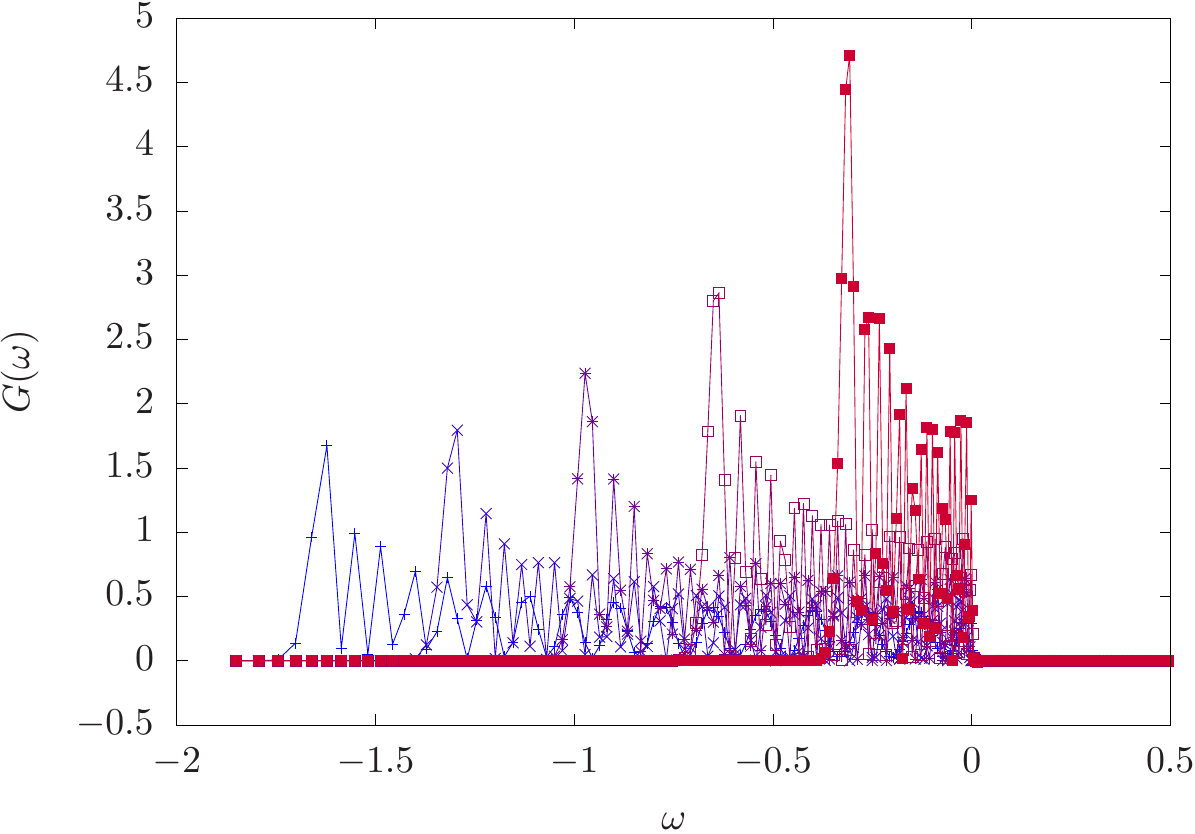}
%\rput(0,5.5){(b)}
\includegraphics[width=8.5cm]{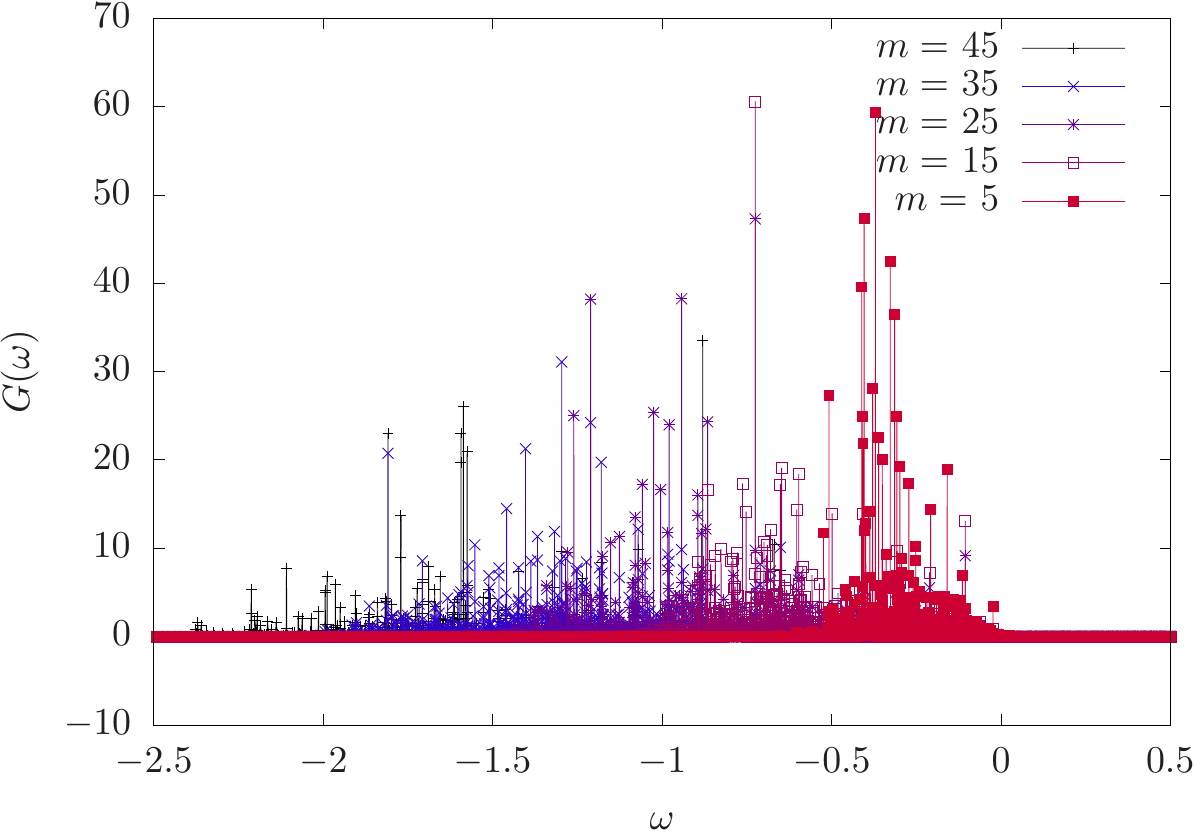}
\caption{(a) Raw response function in the frequency domain for a 1D
  system prior to Gaussian convolution as in
  \eqref{eq:gauss_conv}. The system size is $L_x=500$, and the color
  (and point-type) denoting the distance to the horizon are the same
  as in Fig. \ref{fig:unruh_1d} a. (b) Raw response function in the
  frequency domain for a 2D system prior to Gaussian convolution as in
  \eqref{eq:gauss_conv} and the frequency rescaling described in
  Sect. \ref{sect:signature}. The system size, $L_x=500$, and the
  color code (and point-type) denoting the distance to the horizon are
  the same as in Fig. \ref{fig:unruh_2d} a.}
\label{fig:unruh_1d_raw}
\end{figure}

\subsection{Relation to the Riemann conjecture}

The Dirac Hamiltonian is of interest in very different areas, not
only of physics, but also of mathematics. Indeed, fermionic models
are regularly used as mathematical tools in differential geometry
and analytic number theory. For instance, by studying the number
of non-trivial solutions of a Dirac operator in a given manifold,
it is possible to determine the topological properties of the
manifold itself as proved by the celebrated Atiyah-Patodi-Singer
index theorem \cite{Melrose.93}.

The Dirac Hamiltonian in Rindler spacetime considered in this
paper and its non-spinorial 1D equivalent $H=\sqrt{x} p \sqrt{x}$
provides a handle for proving the Riemann conjecture, which is one
of the most famous and relevant open problems in mathematics.
Riemann conjectured that the non-trivial zeroes of the Riemann
zeta function $\zeta(s)$ in the complex plane all have real part
$1/2$ \cite{Edwards.74}. One of the established routes towards
proving this conjecture --the Hilbert-Polya route-- is specially
interesting for physicists as it attempts the construction of a
Hermitian operator whose eigenvalues are the imaginary parts of
the non-trivial Riemann zeroes. In physics, natural occurrences of
Hermitian operators are, of course, quantum Hamiltonians
\cite{Schumayer.RMP.11}. In 1999, Berry and Keating proposed the
$H=xp$ Hamiltonian and showed how the statistical behavior of its
eigenvalues corresponded to the statistical average behavior of
the imaginary parts of the non-trivial Riemann zeroes
\cite{BerryKeating.99}. In fact, the classical Hamiltonian $H=xp$
must be supplemented with a quantization prescription. The most
natural one is

\begin{equation}
H=\sqrt{x}p\sqrt{x}=-i(\sqrt{x} \pl_x \sqrt{x}) = -i (x\pl_x +1/2).
\label{ham.xpx}
\end{equation}
i.e., the 1D version of the Dirac Hamiltonian in Rindler spacetime. The
discovery of this Berry-Keating Hamiltonian led to a series of
attempts to extend the model in several directions
\cite{Sierra.05,Sierra.11}, including a recent spinorial extension,
which is Eq. \eqref{diracEq.xpx} \cite{Gupta.13}.

\begin{figure}[h!]
%\rput(0,5.5){(a)}
\includegraphics[width=8.5cm]{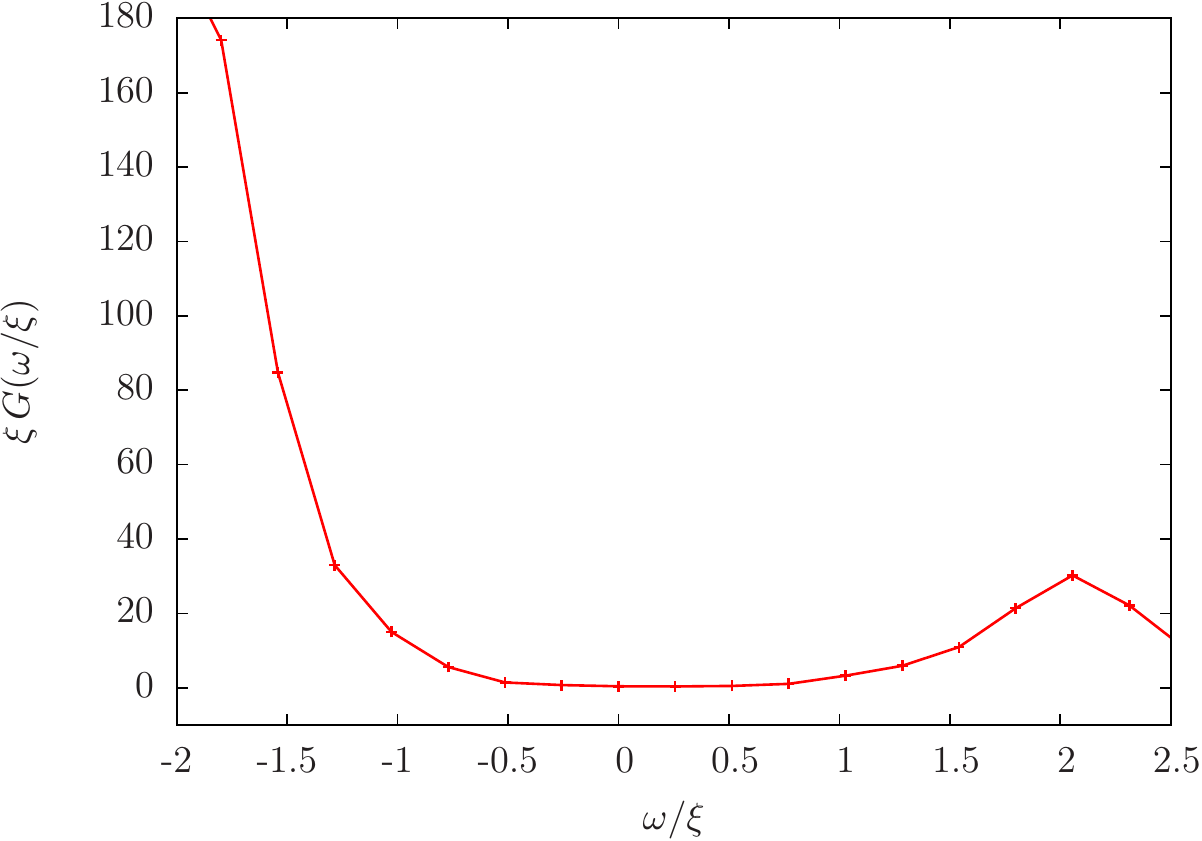}
%\rput(0,5.5){(b)}
\caption{Wightman response function for a 2D thermal gas of Dirac
  fermions in Minkowski space. As expected from the behavior of the
  density of states of the 2D Fermi gas in a $\pi-$ flux lattice, the
  response has a maximum at positive frequency $\omega$. The response
  above coincides with the one obtained in Rindler spacetime in Fig.
  \ref{fig:unruh_thermal} at very large distance from the horizon.}
\label{fig:2d_mink_gas}
\end{figure}

%%% new part
\subsection{Response function for a thermal gas and Gaussian
convolution}\label{sect:A3}

In Sect. \ref{sec:Unruh_effect}, we have derived the expression for
the Wightman response function in the frequency domain, $G(\omega)$,
for the ideal case of fermionic atoms at zero temperature,
\eqref{detector.response.full}.  For a realistic gas at finite
temperature $T$ considered at the end of Sect. \ref{sect:signature},
the Minkowski vacuum, $\ket|0_M>$, appearing in
\eqref{detector.response.full}, has to be replaced by the thermal
mixed state $\rho_M(T)$

\begin{equation}
  \rho_M (T)= \sum_k \frac 1{1+ \exp[\hbar\omega^M_k/k_B T]}
  b_k^\dagger \ket|\Omega> \bra<\Omega|b_k .
\end{equation}
It follows that the  response function for a thermal gas is

\begin{align}
G_{x_0} (\omega) &= \Tr [\rho_M(T) c_{x_0}^\dagger c_{x_0}]\cr
                 &= \sum_{q,q'} \delta(\omega-\omega_q^R) \bar
R_{qx_0} R_{q'x_0} C_{qq'},
\end{align}
where

\begin{equation}
C_{qq'}=\sum_k \bar U_{qk} U_{q'k} \frac 1{1+\exp[\hbar\omega^M_k/k_BT]}.
\end{equation}
Note that for $T\to 0$, $C_{qq'}\to \sum_{\omega^M_k<0}\bar U_{qk} U_{q'k}$, and one
recovers the zero temperature response function
\eqref{detector.response.full}. As defined in the main text, eqs.
\eqref{bogoliubov.transf} and \eqref{rindler.minkowski}, the unitary
matrices $U_{qk}$ and $R_{qx}$ are determined from the single-particle
modes of Dirac Hamiltonian in Minkowski and Rindler spacetime.

In Sect. \ref{sec:Unruh_effect}, in order to smear out lattice
artifacts, we have considered a convolution of the response function
with a Gaussian. In fact, as explained in Sect.
\ref{sect:implementation}, such convolution is what is really detected
by one-particle excitation spectroscopy. The Gaussian convolution
consists of the following. By defining

\begin{equation}
F(x-x_0)=\frac 1{\sqrt{2\pi} \sigma} \exp[-(x-x_0)^2/2\sigma^2],
\end{equation}
the convoluted response function reads

\begin{equation}
G^F_{x_0}(\omega)=\sum_{q,q'} \delta(\omega-\omega_q^R)
\sum_x  F(x-x_0) \bar R_{qx} R_{q'x} C_{qq'}, \label{eq:gauss_conv}
\end{equation}
where $x$'s are the abscissae of the lattice sites, $x=m$ (the lattice
space is taken to be one for convenience).  The response functions
presented in Fig. 2-4 are obtained by taking $\sigma=2$.


\begin{thebibliography}{10}
\bibitem{footnote1}The required TEM$_{10}$ Hermite-Gauss mode could be
  obtained using a phase plate, a spatial light modulator or an
  optical cavity.
\end{thebibliography}
\end{document}